

\input harvmac

\def\Appendix#1{\global\meqno=1\global\subsecno=0\xdef\secsym{\hbox{A.}}
\bigbreak\bigskip\noindent{\bf Appendix. #1}\message{(Appendix.#1)}
\writetoca{Appendix {Appendix.} {#1}}\par\nobreak\medskip\nobreak}

\def\etal{{\it et al.}}
\def\lsim{\mathrel{\rlap{\lower4pt\hbox{\hskip1pt$\sim$}}
    \raise1pt\hbox{$<$}}}         
\def\gsim{\mathrel{\rlap{\lower4pt\hbox{\hskip1pt$\sim$}}
    \raise1pt\hbox{$>$}}}         

\def\rhs{right hand side}

\def\pslash{p\!\!\!/}
\def\kslash{k\!\!\!/}

\def\qslash{q\!\!\!/}
\def\p2inf{\mathrel{\mathop{\sim}\limits_{\scriptscriptstyle
{p^2 \rightarrow \infty }}}}
\def\k2inf{\mathrel{\mathop{\sim}\limits_{\scriptscriptstyle
{k^2 \rightarrow \infty }}}}
\def\x2inf{\mathrel{\mathop{\sim}\limits_{\scriptscriptstyle
{x \rightarrow \infty }}}}
\def\Lam2inf{\mathrel{\mathop{\sim}\limits_{\scriptscriptstyle
{\Lambda \rightarrow \infty }}}}
\def\Q2inf{\mathrel{\mathop{\sim}\limits_{\scriptscriptstyle
{Q^2 \rightarrow \infty }}}}

\def\frac#1#2{{{#1}\over {#2}}}
\def\smallfrac#1#2{\hbox{$\frac{#1}{#2}$}}
\def\half{\smallfrac{1}{2}}
\def\third{\smallfrac{1}{3}}
\def\quarter{\smallfrac{1}{4}}

\def\gev{{\rm GeV}}

\catcode`@=11 
\def\slash#1{\mathord{\mathpalette\c@ncel#1}}
 \def\c@ncel#1#2{\ooalign{$\hfil#1\mkern1mu/\hfil$\crcr$#1#2$}}
\def\lsim{\mathrel{\mathpalette\@versim<}}
\def\gsim{\mathrel{\mathpalette\@versim>}}
 \def\@versim#1#2{\lower0.2ex\vbox{\baselineskip\z@skip\lineskip\z@skip
       \lineskiplimit\z@\ialign{$\m@th#1\hfil##$\crcr#2\crcr\sim\crcr}}}
\catcode`@=12 
\def\twiddles#1{\mathrel{\mathop{\sim}\limits_
                        {\scriptscriptstyle {#1\rightarrow \infty }}}}
\def\matele#1#2#3{\langle {#1} \vert {#2} \vert {#3} \rangle }
\def\PR{{\it Phys.~Rev.~}}
\def\PRL{{\it Phys.~Rev.~Lett.~}}
\def\NP{{\it Nucl.~Phys.~}}
\def\PL{{\it Phys.~Lett.~}}
\def\vol#1{{\bf #1}}
\def\vpy#1#2#3{\vol{#1} (#3) #2}

\noblackbox
\pageno=0
\nopagenumbers
\tolerance=10000
\hfuzz=5pt
\line{\hfill OUTP-93-18P}
\line{\hfill   DFTT 9/93}
\line{\hfill   hep-th/9308322}
\line{\hfill August 1993}
\vskip 10pt
\centerline{\bf ANOMALOUS EVOLUTION OF THE GOTTFRIED SUM}
\vskip 36pt\centerline{Richard D.~Ball\footnote{${}^\star$}{Address
from October 1993 (RDB) and December 1993 (SF): Theory Division, CERN, CH--1211
Gen\`eve 23, Switzerland.}}
\vskip 12pt
\centerline{\it Theoretical Physics, University of Oxford}
\centerline{\it 1 Keble Road, Oxford OX1 3NP, U.K.}
\vskip 10pt
\centerline{ and }
\vskip 10pt
\centerline{Stefano Forte${}^\star$}
\vskip 12pt
\centerline{\it I.N.F.N., Sezione di Torino}
\centerline{\it via P.~Giuria 1, I--10125 Torino, Italy}
\vskip 1.4in
{\narrower\baselineskip 10pt
\centerline{\bf Abstract}
\medskip
We discuss nonperturbative QCD evolution of nonsinglet
nucleon structure functions, with particular application to the
Gottfried sum. We show that the coupling of the
quark partons to bound state mesons leads to nonperturbative
contributions to the Altarelli--Parisi equations which, due to the
axial anomaly, result in a strong scale dependence
of nonsinglet structure functions for values of $Q^2$ around the
nucleon mass scale. We compute specifically the evolution of the first
moment of the quark distribution, and find that it is sufficient to
explain recent experimental data which indicate a violation of the
Gottfried sum rule.
}

\vskip 0.8in
\centerline{Submitted to: {\it Nuclear Physics B}}
\vfill
\eject
\footline={\hss\tenrm\folio\hss}

\newsec{Violation of the Gottfried Sum Rule}

In the wake of the furore over the EMC measurement \ref\emc{J.~Ashman
et al., \PL\vpy{B206}{364}{1988}; \NP\vpy{B328}{1}{1990}.}
of quark polarized structure
functions, experimental results from the NMC\ref\nmc{P.~Amaudruz \etal,
\PRL\vpy{66}{560}{1991}.}
for the Gottfried sum again appear to be in
contradiction with theoretical expectations. More specifically, the NMC
measure the ratio of the cross--sections for unpolarized deep--inelastic
scattering from deuterium and hydrogen targets. The ratio of structure
functions ${F_2^d(x,Q^2)\big{/}F_2^p(x,Q^2)}$ is extracted in the range
$0.004\le x\le0.8$ at $Q^2=4\>\gev ^2$. Extrapolating to small $x$ using Regge
behaviour, and assuming further that the neutron structure function is
just $F_2^n=2F_2^d-F_2^p$ allows a determination of the Gottfried
sum, defined as
\eqn\gsum
{S_G\equiv\int_0^1\!{dx\over x}\,\left[F_2^p(x)-F_2^n(x)\right],}
with the result
\eqn\esum
{S_G=0.24\pm0.016\qquad\hbox{\rm at}\qquad Q^2=4\gev ^2.}
This result (unlike previous less precise experimental
determinations which had a consistent central value but larger errors)
is in striking contradiction with the Gottfried sum rule, which
would have $S_G=\third $ \ref\gott{K.~Gottfried, \PRL\vpy{18}{1174}{1967}
\semi see also e.g.
F.~Close ``An Introduction to Quarks and Partons'' (Academic,
London, 1979)
or E.~Leader and E.~Predazzi ``Gauge Theories and the New Physics''
(Cambridge University, Cambridge, 1982).}.

Both of the assumptions used by the NMC to arrive at the result \esum\ may be
questioned: on the one hand, if Regge behaviour is assumed to set in only
at very small $x$ ($\sim 10^{-3}$ or less) it is not difficult to
increase $S_G$
\ref\msres{A.~D.~Martin, W.~J.~Stirling and R.~G.~Roberts,
\PL\vpy{B252}{653}{1990}\semi
S.~D.~Ellis and W.~J.~Stirling, \PL \vpy{B256}{258}{1991}.},
while on the other if nuclear shadowing is taken into account
the neutron structure function is certainly larger than the difference between
deuterium and proton structure functions, thus leading to a yet smaller value
of $S_G$, perhaps\ref\zbk{V.~R.~Zoller, \PL \vpy{B279}{145}{1992}\semi
B.~Bade\l ek and J.~Kwieci\'nski, \NP\vpy{B370}{278}{1991}\semi
W.~Melnitchouk and A.~W.~Thomas, \PR\vpy{D47}{3783}{1993}.}
around or even below $S_G\sim 0.2$.
The settling of these issues rests ultimately with experiment, and
in particular with the availability of better
data in the small-$x$ region. However, if we accept that the result \esum\
is correct, perhaps with slightly underestimated error, we must conclude that
the Gottfried sum rule does not hold at the scale of the NMC measurement.

Let us therefore review how this sum rule is derived.
In the parton model, $F_2(x)$ is related to the sum of parton densities,
weighted by the electric charges squared (see e.g.\ref\alt
{G.~Altarelli, {\it Phys. Rep.} \vpy{81}{1}{1982}.}):
\eqn\fandq
{F_2(x)=x\sum_ie^2_i\left(q_i(x)+\bar q_i(x)\right),}
where $q_i(x)$ ($\bar q_i(x)$) is the quark (antiquark) density of
flavor $i$. This identification is exact (by definition) to all orders in the
QCD parton model. Although the first moments of $q_i(x)$ and
$\bar q_i(x)$
diverge because of their small-$x$ behaviour, this divergence
is expected to cancel in nonsinglet combinations of quark distributions (such
as in the Gottfried sum), on the basis
the small-$x$ behaviour of $q_i(x)$ expected
from Regge theory\ref\smallx{
B.~Bade\l ek et al., {\it Rev. Mod. Phys.} {\bf 64} (1992) 927.}.

It follows that the Gottfried sum measures the difference
between the sums of the square charges of partons
(quarks {\it plus} antiquarks) in proton and
neutron. Assuming exact isospin symmetry
\eqn\fandn
{S_G=\int_0^1 \!dx\; (e_u^2-e_d^2)
\big(q_u(x)-q_d(x)+\bar{q}_u(x)-\bar{q}_d(x)\big)
=\third (q_1+\bar{q}_1),}
where $q_1$ and $\bar{q}_1$ are the first moments of the nonsinglet
quark distributions
\eqn\qdef
{q(x)\equiv q_u(x)-q_d(x),\qquad \bar{q}(x)\equiv\bar{q}_u(x)-\bar{q}_d(x).}
On the other hand, one may derive an exact sum rule (the Adler sum rule)
for the charge-conjugation odd combination of first moments
which counts quarks {\it minus} antiquarks.  By charge conservation
this is fixed:
\eqn\sadler
{S_A\equiv q_1-\bar{q}_1=1.}
Only valence quarks contribute to the Adler sum rule, since (by
definition) $q_1^{\rm sea}=\bar{q}_1^{\rm sea}$, while the Gottfried
sum receives contributions from both valence and sea quarks. However
if we assume further that the quark--antiquark sea is isotopically
neutral, then $\bar q_1=\bar{q}_1^{\rm sea}=0$, and we find
(using \sadler\ in \fandn) the naive result $S_G=\third$ alluded to above.

The NMC result \esum\ thus suggests that the light flavor
content of the nucleon is very different from the standard
folklore\ref\exppart{See e.g. T.~Sloan, G.~Smadja and R.~Voss,
{\it Phys. Rep.} \vpy{162}{45}{1988}.}
of an asymmetric valence quark content, plus sea quark, antiquark and
gluon distributions which respect all possible symmetries. In
particular, a detailed analysis\ref\mygott
{S.~Forte, {\it Phys Rev.} \vpy{D47}{1842}{1993}.} indicates that
isospin violations (such that the total number of sea quark--antiquark
pairs in the proton and in the neutron are not exactly the same)
cannot by themselves account for the discrepancy; the value \esum\
requires a large
violation of the SU(2) symmetry of the quark--antiquark sea,
such that the
number of up quark--antiquark pairs in each nucleon
must be different from the number of down pairs. More precisely, the NMC
data require\mygott\ the isospin violation
to be $\lsim 5\%$, and the sea asymmetry to be opposite in sign to the valence
asymmetry, and about 30\% of its magnitude. Thus (recalling eqn.\sadler)
$q_1^{\rm sea}+\bar{q}_1^{\rm sea}\sim -0.3$, and not zero as naively assumed.

This conclusion is not only unsettling from the phenomenological point of view,
given that the naive assumption of a symmetric sea
is embodied in all phenomenological parameterizations of parton distributions
which are used in analyzing deep--inelastic scattering
\nref\MRS{A.~D.~Martin, W.~J.~Stirling and R.~G.~Roberts,
\PR\vpy{D47}{867}{1993}; {\it Phys.~Lett.} {\bf B306} (1993) 145;
Rutherford preprints RAL-93-014 and RAL-93-027(1993)\semi
J.~Botts et al., {\it Phys. Lett.} {\bf B304} (1993) 159.}
data\foot{Recently new  parameterizations have appeared,
which incorporate the NMC asymmetry\MRS.},  but is also
puzzling from the theoretical viewpoint. Indeed, one might expect the
quark--antiquark sea probed in deep--inelastic scattering to be mostly produced
through QCD evolution. But the scale dependence
of $S_G$ as computed perturbatively at two loops in
QCD\ref\rs{D.~A.~Ross and C.~T.~Sachrajda, \NP\vpy{B149}{497}{1979}.}
is negligibly weak\foot
{Though not zero, as seems to be sometimes believed
(see for example ref.\exppart).},
leading to a variation of $S_G$ by less than
1\% in the perturbative region. Furthermore, the same computation\rs\
shows explicitly that the naive explanation\ref
\pauli{A.~Signal, A.~W.~Schreiber and A.~W.~Thomas,
{\it Mod. Phys. Lett.} \vpy{A6}{271}{1991}.} of the sea asymmetry
as a consequence of the
Pauli principle (according to which the emission of sea up pairs in the
proton should be reduced due to the up excess in the valence component
\ref\FF{R.~P.~Feynman and R.~D.~Field, \PR\vpy{D15}{2590}{1977}.})
is incorrect: due to subtle interference effects the Pauli
antisymmetrization actually leads to a slight perturbative enhancement
of the sea flavor asymmetry with the same sign as the valence asymmetry.

It has been suggested\nref\thomas
{A.~W.~Thomas, \NP\vpy{A532}{177}{1991}.}\nref\kum
{S.~Kumano, \PR\vpy{D43}{59}{1991} \PR\vpy{D43}{3067}{1991}\semi
S.~Kumano and J.~T.~Londergan, \PR\vpy{D44}{717}{1991}.}
\refs{\pauli,\thomas,\kum}
that the effect may instead be related to the presence of a pion cloud
in the nucleon wave function.
Indeed, if one assumes that some portion of the sea is produced through
transitions where a pion is radiated by a nucleon, then
a proton would seem to favor the transition where a
$\pi^+$  is created, namely $p\to n+\pi^+$, over that where a $\pi^-$ is
created, namely $p\to \Delta^{++}+\pi^-$
(and conversely for a neutron), because of the nucleon-$\Delta$
mass difference. This, spelling out the quark content, is seen
to favor the production of $d\bar d$ pairs over $u\bar u$ pairs, thus
producing a flavor asymmetry with the right sign.
The observed effect
may also be accommodated rather easily in most
of the various effective models of the nucleon; examples are the
Skyrme model\nref\wh{H.~Walliser and G.~Holzwarth,
\PL\vpy{B302}{377}{1993}.}\refs{\mygott,\wh}, chiral soliton models
\ref\waka{M.~Wakamatsu, \PR\vpy{D44}{R2631}{1991};
\PR\vpy{D46}{3762}{1992}.}, and various bag models\nref\thom{
A.~W.~Schreiber et al., \PR\vpy{D45}{3069}{1992}.}\refs{\mygott,\thom}.

The problem with this kind of explanation is not only that it is
difficult to make quantitative and model--independent predictions,
but that the connection between the models (which are presumably
only valid at some unknown but low energy scale) and the parton
distributions measured by the NMC is
not at all clear.
Furthermore, even accepting that pion effects
are somehow responsible for the observed asymmetry, it is not
enough\mygott\ to assume that the $\pi^+$ and $\pi^-$ clouds in a
nucleon are unequal, since each pion gives a vanishing
contribution to the flavor asymmetry of the quark sea. Rather, one
has to make a separate dynamical assumption on the pion production mechanism,
in order to obtain an anticorrelation between valence and sea
asymmetries. Such a mechanism has been presented in ref.\ref
\ehq{E.~J.~Eichten, I.~Hinchliffe and C.~Quigg, \PR\vpy{D45}{2269}
{1992}.}, where it is shown that if (part of) the sea is produced by
dissociation of a quark into a quark plus a pion, then the breaking of the
axial U(1) symmetry in the pseudoscalar spectrum favors flavor--changing
transitions over flavor--preserving ones. A calculation \ehq\
based on the chiral quark model of
ref.\ref\gm{H.~Georgi and A.~Manohar, \NP\vpy{B234}{189}{1984}.}
seems to support this idea.

It is interesting to compare this situation with a similar instance of
the failure of naive expectations, namely that of the Ellis--Jaffe sum
as measured by the EMC\emc\ (for reviews see
ref.\ref\spinrev{G.~Altarelli, in ``The Challenging Questions'', Proc. of
the 1989 Erice school, A.~Zichichi, ed. (Plenum, New York, 1990)
\semi G.~G.~Ross in ``Proc. of the 1989 Int. Symp. on Lepton and
Photon Interactions at High Energy'', Stanford, 1989 (World Scientific,
Singapore, 1989)
\semi G.~Veneziano in ``From Symmetries to Strings'', proc. of
the Okubofest, Rochester, 1989, A.~Das, ed. (World Scientific, Singapore,
1990)
\semi J.~Ellis, \NP\vpy{A546}{447c}{1992}.}).
In that case too the first moment of a nucleon structure function
(the polarized proton structure function $g_1^p$) is found to be
considerably smaller than the parton model expectation,
implying that the quark
contribution to the ``proton spin'' (more precisely,
its isosinglet axial charge) is
consistent with zero. 	There, however, perturbative QCD suggests
a partial explanation of the observed effect:
due to the axial anomaly, the isosinglet axial charge evolves, despite
being classically conserved. This evolution may be
interpreted in the parton model by writing the charge as a sum of
a polarized quark contribution (which does not evolve) and an
anomalous gluon contribution (which does).
Whereas the perturbative evolution (which, again, only
occurs at two loops) is too weak to explain why the gluon
contribution is so large,
the observed discrepancy could be explained by a nonperturbative dynamical
mechanism which generates a gluon contribution of the appropriate sign
and magnitude. Such a mechanism would incorporate the same infrared
non--perturbative physics that resolves the U(1) problem by giving
a mass to the flavor singlet pseudoscalar meson (the $\eta_0$)
\ref\uone{G.~'t~Hooft, {\it Phys. Rep.} \vpy{142}{357}{1986}
and ref. therein.}. It could take the form of
infrared vacuum effects (as modelled for example by instantons)
\ref\protfor{S.~Forte, \NP\vpy{B331}{1}{1990}\semi
S.~Forte and E.~V.~Shuryak, \NP\vpy{B357}{153}{1991}.}
which reduce the starting value of the charge, or a non--perturbatively
induced scale dependence of the isosinglet axial charge
\ref\protball{R.~D.~Ball, \PL\vpy{B266}{473}{1991}.}.

Here we will show that the anomaly also affects the non--perturbative
evolution of the Gottfried sum, albeit in a rather more subtle way.
We compute the non--perturbative scale dependence of
nonsinglet quark structure functions by including bound
state emission, thereby generalizing the usual Altarelli--Parisi
evolution equations. We find that when the axial anomaly is taken
into account through the generation of a mass for the $\eta_0$
(as suggested in ref.\ehq), then
a flavor asymmetry in the quark sea is generated dynamically
 and the Gottfried sum
acquires a rather strong scale dependence over intermediate scales.
At very large and very small scales
this dependence flattens out, so that the two loop perturbative evolution
eventually takes over at large $Q^2$, whereas a constant ``quark model'' value
is attained in the infrared region. We find that if this is
identified with the naive expectation $S_G=1/3$, then the experimental
value of $S_G$, eqn.\esum, can be easily reproduced.

The paper is organized as follows: in section 2 we review the
general theory of the $Q^2$ dependence of the
nonsinglet structure function $F_2$, both from the point of view of the
operator--product expansion, and the QCD parton model; in section 3 we
generalize the latter approach to include non--perturbative coupling to
mesonic bound states, we describe the determination of the
nonperturbative contribution to the anomalous dimensions,
and we discuss how this can lead to flavor symmetry breaking effects due
to the U(1) anomaly; in section 4 we compute the non--perturbative
contribution to the appropriate splitting functions
by explicit determination of the relevant cross--sections;
in section 5 we determine the $Q^2$ dependence of the
first moment of the quark distribution, hence of the Gottfried sum,
and compare our results with the NMC data; and finally in
section 6 we summarize our results and discuss how this approach may
be developed into a more general technique for the computation of
nonperturbative contributions to structure function evolution.

\newsec{Evolution of Nonsinglet Quark Distributions}

Because the structure functions $F_2(x)$ are proportional \fandq\ to the
quark distributions $q_i(x)$, the problem of the determination
of their $Q^2$ dependence reduces to the computation of the
QCD evolution of the latter. Since we are only interested
in the difference $F_2^p(x)-F_2^n(x)$, we need only discuss the
evolution of the nonsinglet distributions $q(x)$ defined in \qdef,
although most of what follows would apply equally to other nonsinglet
distributions.

We begin with a discussion of the evolution as dictated by the
operator product expansion\rs.
This is not straightforward, because the (electromagnetic)
structure functions $F_2$ are even under charge conjugation, as eqn.\fandq\
demonstrates explicitly. It follows that at leading twist only the even
moments of $F_2$ can be expressed as the matrix element of a local
operator, since the leading--twist contribution to the \hbox{$N$-th} moment
has spin $N$ hence charge conjugation ${\cal C}=(-)^N$. As shown in ref.\rs,
however, this does not imply that it is impossible to give a meaning to odd
moments of $F_2^p$ or their evolution, as is sometimes claimed\ref
\mano{A.~V.~Manohar, Lectures at Lake Louise Winter Inst.,
Canada, 1992, San Diego preprint UCSD/PTH 92-10 (1992).}.

Indeed, consider the combinations
\eqn\qpm
{q^{\pm} (x;t)= q(x;t)\pm \bar q(x;t),}
where we have indicated explicitly the dependence on the scale
$t\equiv\ln Q^2\big/\mu ^2$.
Since $q^{\pm}$ are eigenstates of charge conjugation, they evolve
independently. Even moments of $q^+$ and odd moments $q^-$ have
the scale dependence
\eqn\evolold
{\eqalign{q_N^+(t)&={\Delta}_N(t,t_0)\,q_N^+(t_0),
\qquad N\;\hbox{even},\cr
q_N^-(t)&={\Delta}_N(t,t_0)\,q_N^-(t_0),
\qquad N\;\hbox{odd},\cr}}
where $q_N^{\pm}(t)\equiv\int_0^1 dx\>x^{N-1}q^{\pm}(x;t)$ is the
$N$-th moment of $q^{\pm}(x;t)$, $t_0\equiv\ln Q_0^2/\mu^2$, and the
scaling factor ${\Delta}_N(t,t_0)$ is given in terms of the
anomalous dimensions $\gamma_N(t)$ of the \hbox{twist-2},
\hbox{spin-$N$} nonsinglet local operators by
\eqn\de
{\ln {\Delta}_N(t,t_0)=\int_{t_0}^{t}\!\gamma_N(t')\, dt'.}
If we let
\eqn\gapm
{\gamma_N=\gamma_N^{qq}+(-)^N \gamma^{q\bar q}_N}
and assume that both $\gamma_N^{qq}$ and $\gamma^{q\bar q}_N$ admit
an analytic continuation for all values of $N$ (as may be verified
explicitly in perturbation theory\rs),
it follows that the evolution equations
\evolold\ also admit a continuation for all values of $N$, such that
$q_N^{\pm}$ evolve multiplicatively according to
\eqn\evol
{q_N^{\pm}(t)={\Delta}_N^{\pm}(t,t_0)q_N^{\pm}(t_0),}
where ${\Delta}_N^{\pm}(t,t_0)$ are defined in terms of the
anomalous dimensions
\eqn\gapmall
{\gamma_N^{\pm}(t)=\gamma_N^{qq}(t)\pm \gamma^{q\bar q}_N(t).}
In particular, the Gottfried sum evolves multiplicatively as
\eqn\gottevol
{S_G(Q^2)={\Delta}_1^+\left(t,t_0\right)S_G(Q_0^2).}

This somewhat formal result becomes more transparent in the parton
model, where the evolution of the quark distributions
is found by using the Altarelli--Parisi equations\alt
\eqn\apeq
{\eqalign{{d\over dt} q_i &=\sum_j  {\cal P}_{q_iq_j}\otimes q_j+
\sum_j {\cal P}_{q_i\bar q_j}\otimes \bar  q_j+{\cal P}_{q_i G}\otimes
G, \cr
{d\over dt} \bar{q}_i &=\sum_j  {\cal P}_{\bar{q}_i\bar{q}_j}\otimes \bar{q}_j+
\sum_j  {\cal P}_{\bar{q}_i q_j}\otimes q_j+
 {\cal P}_{\bar{q}_i G}\otimes G,\cr}}
where ${\cal P}$ are splitting functions, the sum runs over flavors,
and $ {\cal P}\otimes q\equiv\int_x^1\!\frac{dy}{y}\,
{\cal P}\big(\frac{x}{y}\big)q(y)$.
Since, by charge conjugation,
${\cal P}_{q_iq_j}={\cal P}_{\bar q_i\bar q_j}$ and
${\cal P}_{q_i\bar q_j}={\cal P}_{\bar q_i q_j}$, if we assume all
relevant flavors to be massless there are only four distinct
quark--quark or quark--antiquark
splitting functions, namely ${\cal P}^{D}_{qq},{\cal P}^{D}_{q\bar q},
{\cal P}^{ND}_{qq}$ and ${\cal P}^{ND}_{q\bar q}$, where
${\cal P}^D$ (${\cal P}^{ND}$) is any ${\cal P}_{ij}$ such that $i=j$
($i\not =j$). The evolution of the nonsinglet quark and antiquark
distributions \qdef\ is then simply given by
\eqn\nsingq
{\eqalign{&{d\over dt} q=  {\cal Q}_{qq}\otimes q
+{\cal Q}_{q\bar q}\otimes \bar q,\cr
&{d\over dt} \bar q=  {\cal Q}_{qq}\otimes \bar q
+  {\cal Q}_{q\bar q} \otimes q;\cr
&\quad {\cal Q}={\cal P}^D-{\cal P}^{ND}.
}}
This shows immediately that the eigenstates of the Altarelli--Parisi evolution
are $q^\pm$, eqn.\qpm, since
\eqn\evolpm
{{d\over dt}{q^\pm}=
\left({\cal Q}_{qq}\pm {\cal Q}_{q\bar q}\right)\otimes {q^\pm} .}
Taking moments of this equation and
comparing the result with \evol -\gapmall\
it is clear that the anomalous dimensions
$\gamma_N^\pm$ \gapm\ are just the $N$-th moments of the
evolution kernel on the \rhs\ of eqn.\evolpm, whereas
$\gamma_N^{qq}$ ($\gamma_N^{q\bar q}$) are the $N$-th moments of
${\cal Q}_{qq}$ (${\cal Q}_{q\bar q}$); this may be confirmed
in perturbation theory by explicit computation. In particular, the
nonsinglet evolution of $F_2$ (and hence that of the Gottfried sum) is
found by taking moments of ${\cal Q}_{qq}+ {\cal Q}_{q\bar q}$.

In the parton model several useful perturbative properties of the
anomalous dimensions $\gamma_N^\pm$ become immediately apparent.
Firstly, at one loop
there is only one diagram which may contribute to nonsinglet
evolution, namely gluon radiation by a quark line. Hence all the
splitting functions ${\cal P}_{q\bar q}$ vanish.
On the other hand, at two loops the splitting functions
${\cal P}_{q\bar q}^D $  and ${\cal P}_{q\bar q}^{ND}$ do not vanish,
since a quark may radiate a gluon which in turns radiates a $q\bar q$
pair (see fig.~1). Furthermore the splitting
functions ${\cal P}_{q\bar q}^D $  and ${\cal P}_{q\bar q}^{ND}$
are in general different already at two loops: the cross
sections for flavor--diagonal and flavor--non-diagonal production differ
since in the flavor diagonal case there are two identical particles
in the final state, which must be antisymmetrized.
It follows that all the $\gamma_N^{q\bar q}$ start at two
loops. Furthermore, specializing to the first moment, charge conservation
implies that the first moment of $q^-$ must be conserved (and thus
that the Adler sum \sadler\ does not evolve). It follows that
$\gamma_1^{qq}=\gamma_1^{q\bar q}$, and in particular that they both
vanish at one loop.

In conclusion, we see that in perturbative QCD the evolution of the
Gottfried sum is governed by the anomalous dimension
$\gamma^+_1=2\gamma_1^{qq}$,
which is the first moment of the splitting
functions for diagonal minus non--diagonal emission. This quantity starts
at two loops, thus leading to a perturbative contribution
to the evolution factor of the form
\eqn\evolpert
{({\Delta}^+_1)_{\rm pert}(t,t_0)=
1+{\gamma_1^{+(2)}\over {b^{(1)}}}
\big(\alpha_s(t)-\alpha_s(t_0)\big)+\cdots}
where $\gamma_1^{+(2)}$ is the two--loop coefficient in
$\gamma_1^+$, and $b^{(1)}$ is the one--loop coefficient in the
beta function. Explicit computation shows that\rs\
${\gamma_1^{+(2)}\!\big/ b^{(1)}}\simeq 0.01$; as mentioned in the
introduction the perturbative evolution is thus entirely negligible
unless $\alpha _s$ is so large that perturbation theory is useless.

This result was derived in the flavor--symmetric, vanishing quark mass limit.
In principle, it may be corrected by two different kinds of effects.
Firstly, if current quark masses are taken into account, then the
simple evolution equation \evol\ is corrected by flavor--violating terms.
These are usually assumed to be negligible, in that mass effects should be
suppressed by powers of $m^2/Q^2$, although they have never been
determined explicitly. If one also admits the possibility of
isospin violation, then $S_G$ also contains a term proportional to the
proton--neutron difference of singlet structure functions, and the
evolution of these must also be taken into account; $S_G$ will then
no longer evolve multiplicatively.  This effect has been discussed
in ref.\mygott, and is negligible for realistic amounts of isospin
violation. We will neglect both these effects henceforth and assume
isospin symmetry to be exact.

\newsec{Evolution by Bound State Emission}

The physical reason for the smallness of the perturbative evolution
found in the previous section should be readily apparent: if we
were to neglect final--state
antisymmetrization effects, then the interaction which governs the QCD
evolution would be fully symmetric under the $U(N_f)$ flavor group,
and the Wigner--Eckart theorem would imply that
${\cal P}^D={\cal P}^{ND}$. There would then be no nonsinglet evolution,
and the dynamically generated sea would be flavor symmetric, as naively
expected. Final--state antisymmetrization effects spoil this argument,
and provide a (tiny) dynamical violation of the flavor symmetry.

However, a much larger violation appears if we take into account\ehq\
that in QCD the $U(N_f)$ flavor symmetry
is dynamically broken due to the axial
anomaly\uone. In particular, consider diagrams where a quark couples
directly to a $q\bar q$ bound state (fig.~2). We may view these as
generated dynamically from diagrams which have
the structure of the perturbative ones of fig.~1, but where the emitted
(and unobserved) quark--antiquark pair may interact non--perturbatively.
These diagrams will be suppressed at large scales where, due to the
extended nature of the emitted bound states,
they must lead to higher--twist contributions to the evolution
equations --- contributions suppressed
by powers of $1/Q^2$ in the large $Q^2$ limit. However at
low enough scales their contribution to nonsinglet evolution
will probably be much greater than the perturbative diagrams. At very
low scales it is no longer clear that the Altarelli--Parisi evolution
picture is applicable at all. There might nevertheless be an intermediate
region where the Altarelli--Parisi picture applies and the evolution
due to bound--state emission is significant.

First, consider the situation in the chiral limit, with $N_f$ massless quarks.
Then, it is clear that all of the bound state emission diagrams will respect
$U(N_f)$ flavor symmetry, and thus will not contribute to nonsinglet
evolution, with the sole exception of those in which pseudoscalar
mesons are emitted. This is because it is only in the
spectrum of pseudoscalar mesons that
$U(N_f)$ is broken down to $SU(N_f)$ by non--perturbative effects
coupled through the axial anomaly\uone; the $N_f^2-1$ Goldstone
bosons will make a significant contribution to nonsinglet evolution
which is not cancelled off by the contribution of the singlet
pseudoscalar (the $\eta_0$) since this is kinematically suppressed.
Away from the chiral limit, the emission of other mesons will also
begin to contribute, but their contributions will remain relatively
small mainly because the mass splittings are themselves relatively
small, but also because they are in any case kinematically suppressed
when compared to those of the light pseudoscalars.
It follows that if we concentrate on nonsinglet evolution, then of the
infinite tower of diagrams for bound--state emission (fig.~2) only
those corresponding to pseudoscalar meson emission need be considered.

To make this discussion more quantitative consider the inclusion of
pseudoscalar bound states $\Pi^a$ in the evolution equation \apeq.
We now have (omitting gluon contributions for simplicity)
\eqn\appi
{\eqalign{{d\over dt} q_i &=\sum_j  {\cal P}_{q_iq_j}\otimes q_j+
\sum_j  {\cal P}_{q_i\bar q_j}\otimes q_j +
\sum_a {\cal P}_{q_i \Pi^a}\otimes \Pi^a,\cr
{d\over dt} \bar{q}_i &=\sum_j {\cal P}_{\bar{q}_i\bar{q}_j}\otimes \bar{q}_j+
\sum_j {\cal P}_{\bar{q}_i q_j}\otimes q_j+
\sum_a {\cal P}_{\bar{q}_i \Pi^a}\otimes\Pi^a,\cr
{d\over dt} \Pi^a &=\sum_j {\cal P}_{\Pi^a q_j} \otimes q_j +
\sum_j  {\cal P}_{\Pi^a \bar{q}_j} \otimes\bar{q}_j+
\sum_b {\cal P}_{\Pi^a \Pi^b}\otimes\Pi^b,\cr}}
where $\Pi^a(x;t)$ are the distribution functions of the pseudoscalar
mesons (the pion octet and the $\eta^\prime$). The first two equations in
\appi\ express the evolution of the quark and antiquark distribution due to
pseudoscalar emission, whereas the last equation gives the evolution of the
pseudoscalar meson distribution. The former is determined by the splitting
functions ${\cal P}_{q q}$,  which expresses the
probability of  a quark to be emitted by another quark
(with emission of an unobserved  pseudoscalar), and  ${\cal P}_{q \Pi}$,
which expresses the probability of a pseudoscalar to fragment into a
quark and an antiquark  (one of which is observed).
The latter
is determined by the splitting functions ${\cal P}_{\Pi q}$
and ${\cal P}_{\Pi\, \Pi}$, which give the
probability of a pseudoscalar to be emitted by a quark or another pseudoscalar,
respectively.

Isospin and charge conjugation invariance imply that for neutral
pseudoscalars $\Pi^0\equiv(\pi^0,\,\eta,\,\eta^\prime)$
\eqn\pineu
{{\cal P}_{u\Pi^0}={\cal P}_{d\Pi^0}={\cal P}_{\bar u\Pi^0}=
{\cal P}_{\bar d\Pi^0},}
whereas for charged mesons
\eqn\pichar
{\eqalign{
&{\cal P}_{u\pi^+}={\cal P}_{d\pi^-}={\cal P}_{\bar d\pi^+}=
{\cal P}_{\bar u\pi^-}\equiv {\cal P}_{q\pi},\cr
&\quad {\cal P}_{d\pi^+}={\cal P}_{u\pi^-}={\cal P}_{\bar u\pi^+}=
{\cal P}_{\bar d\pi^-}=0.\cr}}
The non--perturbative evolution of the nonsinglet quark
and antiquark distributions \qdef\
and the pion distribution
\eqn\pidef
{\pi(x;t)\equiv\pi^+(x;t)-\pi^-(x;t),}
is thus given by
\eqn\evolqqbarpi
{\eqalign{&{d\over dt} q=  {\cal Q}_{qq} \otimes q
+{\cal Q}_{q\bar q}\otimes \bar q+ {\cal P}_{q\pi}\otimes\pi,\cr
&{d\over dt} \bar q=  {\cal Q}_{qq} \otimes\bar q
+  {\cal Q}_{q\bar q}\otimes q- {\cal P}_{q\pi}\otimes \pi,\cr
&{d\over dt} \pi=  {\cal P}_{\pi q} \otimes (q-\bar{q})
+ {\cal P}_{\pi\pi} \otimes\pi .\cr}}
The combinations $q^{\pm}$ \qpm\ then satisfy
\eqn\evolpmpi
{\eqalign{&{d\over dt} q^+= ({\cal Q}_{qq}+{\cal Q}_{q\bar{q}})\otimes q^+ ,\cr
&{d\over dt} q^-=  ({\cal Q}_{qq}-{\cal Q}_{q\bar{q}})\otimes q^-
+{\cal P}_{q\pi}\otimes2\pi ,\cr
&{d\over dt} \pi=  {\cal P}_{\pi q}\otimes q^-
+  {\cal P}_{\pi\pi}\otimes\pi,\cr}}
where, to first order, only the splitting functions
${\cal P}_{qq}={\cal P}_{\bar{q}\bar{q}}$, ${\cal P}_{q\Pi}$,
${\cal P}_{\bar{q}\Pi}$, ${\cal P}_{\Pi q}$ and ${\cal P}_{\Pi
\bar{q}}$ are nonvanishing\foot{
${\cal P}_{\Pi\Pi}$ begins at second order (just as the usual
perturbative contributions to ${\cal P}_{qq}$ only begin at second
order), while although ${\cal P}_{q\bar{q}}$ may receive nontrivial
contributions from diquark emission, the $U(N_f)$ symmetry of the
diquark is not broken anomalously, and thus they may be ignored for
the same reason as we ignore the contribution of all mesons other
than pseudoscalars to ${\cal P}_{qq}$.}
(and in particular ${\cal Q}_{q\bar{q}}=0$).

To actually compute the nonsinglet evolution due to pseudoscalar meson
emission, we use the probabilistic interpretation\ref\prob{L.~Durand and
W.~Putikka, \PR\vpy{D36}{2840}{1987}\semi
J.~C.~Collins and J.~Qiu, \PR\vpy{D39}{1398}{1989}.}
of the
Altarelli--Parisi equations in which the splitting functions
${\cal P}_{qq}$ are obtained through the same sort of procedure which
leads to the perturbative splitting functions\alt . That is, QCD evolution
is in general due to the fact that in deep inelastic scattering a parton may
radiate another particle (perturbatively a gluon, but for present
purposes a pseudoscalar bound state, $\Pi$) either before or after in is
struck by the virtual photon. The splitting function
${\cal P}_{q_iq_j}(x)$ expresses the probability of the quark parton $q_j$
to carry the fraction $x$ momentum of the quark parton $q_i$ after the
radiation process, and it may be obtained obtained from
the cross--section for Compton--like scattering of a (virtual) photon
$\gamma^*$ before or after emission of a bound state (fig.~3).
This corresponds to an effective resummation of the same ladder diagrams which
lead to the usual Altarelli--Parisi evolution, with the exchanged gluons
replaced by pseudoscalar mesons. These diagrams may be shown to
give the dominant contribution in the leading logarithmic
approximation  in a theory where quarks are coupled to pointlike
pseudoscalar particles in just the same way as for
perturbative QCD, but without the complications arising from gauge
invariance\nref\pscAP{S.~J.~Chang and P.~M.~Fishbane,
\PR\vpy{D2}{1084}{1972}\semi V.~N.~Gribov and L.~N.~Lipatov,
{\it Sov.~Jour.~Nucl.~Phys.~}\vpy{15}{438,~675}{1972}\semi
see also Y.~L.~Dokshitzer, D.~I.~Diakonov and
S.~I.~Troyan, {\it Phys.~Rep.~}\vpy{58}{269}{1980}.}
\nref\CHLStwo{C.~H.~Llewellyn~Smith, in
``Facts and Prospects of Gauge Theories'', proc. of the 1978
Schladming school, P.~Urban, ed. (Springer, Wien, 1978).}
\refs{\pscAP,\CHLStwo}. In section 4 we will discuss the
applicability of this picture to extended pseudoscalars.

The contribution of the emission of a bound state $\Pi$ to the
splitting function ${\cal P}_{q_iq_j}$ is then given by
\eqn\splitX
{\left[{\cal P}_{q_iq_j}(x;t)\right]_\Pi=
\frac{d}{dt}\sigma^{\gamma^*\Pi}_{q_iq_j}(x;t),}
where $\sigma^{\gamma^*\Pi}_{q_iq_j} (x;t)$ is the total cross section for
emission of the state $\Pi$ with quantum numbers corresponding to the
given splitting function, expressed in terms of
the usual scaling variables and integrated over all $k_\perp$.
This cross--section is adimensional, due to the extraction of a scale
factor\alt. Notice that the so--called ``loss''
terms\refs{\alt,\prob},
namely those terms which give the decrease in the probability of finding
a quark with momentum fraction $x$ due to the fact that if such a
quark undergoes a radiation process it is lost
from the observed momentum interval, are absent from eqn.\splitX, which gives
the full splitting function.
This is because when a meson is emitted the quark which is struck by
the virtual photon is never the same as the original one, even if it
has the same flavor.

The non--perturbative contribution to the anomalous dimension
${\gamma}^{q_iq_j}_N$ is then found by taking moments of the
splitting function:
\eqn\npad
{\left[\gamma^{q_iq_j}_N(t)\right]_\Pi
=\int_0^{x_{\rm max}}
\!dx\,x^{N-1}\left[{\cal P}_{q_iq_j} (x;t)\right]_\Pi
=\frac{d}{dt} \int_0^{x_{\rm max}}
\!dx\,x^{N-1}\sigma^{\gamma^*\Pi}_{q_iq_j} (x;t).}
Notice that the upper limit of the $x$-integration is not $1$ but
rather
\eqn\xmeq
{x_{\rm max}={1\over 1+{M^2\over Q^2}}\leq 1,}
where $M$ is the mass of the pseudoscalar bound state. The
cross--section vanishes for $x>x_{\rm max}$. By continuity we also
expect  $\sigma^{\gamma^*\Pi}_{q_iq_j} (x_{\rm max};t)=0$; this
justifies the exchange in the order of differentiation and integration
in the second expression \npad.
The anomalous dimensions $\gamma_N^{q\Pi}$ and $\gamma_N^{\Pi q}$
are similarly determined by taking moments of the splitting functions
${\cal P}_{q\Pi}$ and ${\cal P}_{\Pi q}$ obtained by differentiating
the cross--sections $\sigma_{q\Pi}$ and
$\sigma_{\Pi q}$.

Besides the usual crossing symmetry relations among splitting
functions \alt,  charge conservation as expressed by
the Adler sum rule eqn.\sadler\ imposes a nontrivial consistency
condition on the first moments of the various splitting functions. Combining
\sadler\ with the evolution equation \evolpmpi\ and taking the first
moment gives
\eqn\consa
{\gamma_1^{qq}(t)+2{\gamma}^{q\pi}_1(t)\pi_1(t)=0,}
where $\gamma_1^{qq}$ is the first moment of ${\cal Q}_{qq}$, and
$q_1(t)$  and $\pi_1(t)$ are the first moments of
$q(x;t)$ and $\pi(x;t)$ respectively. If we now assume that there exists
a scale $t_0$ such that $\pi_1(t_0)=0$, the entire pion distribution
$\pi(x;t)$ is generated dynamically through the relevant evolution
equation \evolpmpi. Integrating up the first moment of this evolution,
and using the result in eqn.\consa\ leads to the condition
\eqn\consb
{\gamma_1^{qq}(t)+2{\gamma}^{q\pi}_1(t)\left[{\sigma}_{\pi q}^1(t)
-{\sigma}_{\pi q}^1(t_0)\right]=0,}
where ${\sigma}_{\pi q}^1$ is the first moment of the appropriate
cross--section, so that $\gamma^{\pi q}_1(t)={d\over dt}{\sigma}_{\pi q}^1$.
Now eqn.\consb\ means that  $\gamma_1^{qq}(t_0)=0$; the assumption
that $\pi_1(t)$ is entirely generated by the evolution eqn.\evolpmpi\ is
consistent only if the evolution of $q_1(t)$ flattens at small $t$.
The scale $t_0$ at which the evolution flattens is then the same as that
where $\pi_1(t)$ vanishes.

In this formalism the so--called Sullivan process\ref
\sull{J.~D.~Sullivan, \PR\vpy{D5}{1732}{1972}
\semi A.~W.~Thomas, \PL\vpy{B126}{97}{1983}.},
where the virtual photon scatters inelastically on the quarks in a pion
which is in turn carrying a fraction of the nucleon momentum, is
regarded  as a measurement of the $\Pi^a(x)$ structure functions
through their fragmentation into quark--antiquark pairs, as
described by the splitting function ${\cal P}_{q\Pi}$.
This is the nonperturbative analogue, for pseudoscalar bound
states, of a perturbative measurement of the gluon distribution via
photon--gluon fusion. Because some of the bound states we consider are charged,
there is  also the possibility of a direct coupling of the
virtual photon to these states, corresponding to processes where
the pion does not fragment incoherently, but rather is viewed
as a pointlike effective constituent. Such processes, however,
are suppressed by the square of the pion form factor, evaluated at
the photon's virtuality $Q^2$, and are thus negligible \sull.
Furthermore they are only relevant to the evolution of singlet
structure functions; for nonsinglets, and in particular for the
Gottfried sum, they vanish identically because of isospin.

Eqns \appi-\splitX\ provide us thus with a well--defined framework in
which to calculate nonperturbative contributions to the evolution of
structure functions due to the inclusion of pseudoscalar
mesons as effective constituents in the ladder diagrams which lead to
Altarelli--Parisi evolution.
In order to avoid double counting, the non--perturbative contribution to the
anomalous dimensions determined in this way should not be added
directly to the usual perturbative anomalous dimensions. Rather,
the non--perturbative contribution, while dominant at small $Q^2$,
must fall as a power at large $Q^2$, eventually falling below the
logarithmic perturbative evolution. At around this point (which would
ideally be in a region where both calculations can still be believed)
the non--perturbative and perturbative anomalous dimensions should
be matched together. In the particular case of the Gottfried sum the
perturbative evolution eqn.\evolpert\ is so tiny that it may be
ignored; the only significant evolution is nonperturbative.

We specialize now to the evolution of $q^+(x;t)$.
It is apparent from eqn.\evolpmpi\ that the pion
distribution $\pi(x;t)$ makes no direct contribution
to this evolution. This is to be expected \mygott\ on the basis of
a simple quark counting argument: because quarks
and antiquarks contribute with the same sign to Gottfried sum
eqn.\gsum,\fandq, it follows that any pion always give a vanishing contribution
to $S_G$ eqn.\gsum. The emission of bound states contributes nevertheless
indirectly, by leading to nonperturbative evolution of $q^+(x;t)$. The
evolution of the Gottfried sum discussed in the present approach
thus appears to be due to a mechanism which is physically distinct
from the Sullivan process; indeed, in the present
approach scattering on the pion component of the nucleon does not contribute at
all to $S_G$, contrary to the suggestion in ref.\kum\foot
{Notice however that in some specific low energy models the Sullivan
process is actually related\thomas\ by
the dynamics of the pion radiation mechanism to the distribution of quarks in
the recoil hadron after pion emission, thus it may contribute indirectly to
quark distributions in a way which is analogous to that which we discuss here.
In
these models, however, there is no scale dependence of structure functions and
quark distributions.}.

The non--perturbative evolution of the $N$-th moment of the nonsinglet
quark distributions $q_N^+(x)$ \qdef,\qpm\ is  found by
computing the anomalous dimensions \npad\ for the $u$ and $d$ flavors
and the pseudoscalar mesons $\pi^\pm$, $\pi ^0$, $\eta$
and $\eta '$. For simplicity we
may assume that the latter are pure octet and pure singlet, since the
appropriate mixing angle $\theta _P$ is small
($\sin^2\theta_P\simeq 0.1$). Noting that the probability for
flavor--diagonal emission by a light quark is obtained by adding the
probabilities for the process to go through $\pi^0$, $\eta$ and
$\eta'$ meson emission as dictated by their quark content we find
\eqn\lightev
{\eqalign{&\gamma^{ud}_N=\gamma^{du}_N=
{d\over dt}\sigma^{\gamma^*\pi^+}_N ={d\over dt}\sigma^{\gamma^*\pi^-}_N,\cr
&\gamma^{uu}_N=\gamma^{dd}_N={d\over dt}
\left({\half\sigma^{\gamma^*\pi^0}_N+\smallfrac{1}{6}\sigma^{\gamma^*\eta}_N
+\third\sigma^{\gamma^*\eta '}_N}\right),\cr
&\gamma^{us}_N=\gamma^{su}_N=
{d\over dt}\sigma^{\gamma^*K^+}_N ={d\over dt}\sigma^{\gamma^*K^-}_N,\cr
&\gamma^{ds}_N=\gamma^{sd}_N=
{d\over dt}\sigma^{\gamma^*K^0}_N ={d\over dt}\sigma^{\gamma^*\bar K^0}_N,\cr
&\gamma^{ss}_N={d\over dt}\left({\smallfrac{2}{3}\sigma^{\gamma^*\eta}_N
+\third\sigma^{\gamma^*\eta '}_N}\right),\cr}}
where $\sigma^{\gamma^*\Pi}_N$ is the $N$-th moment (defined as in the
\rhs\ of eqn.\npad) of the cross section for the process of fig.~3, and
we have included for completeness strange quarks and kaon emission.
Using eqn.\lightev\ in the evolution equation
\evolpmpi\ we finally get
\eqn\evolud
{\eqalign{\gamma _N^+ =\gamma_N^{qq}=\gamma_N^{uu}-\gamma_N^{ud}&={d\over dt}
\left(\half\sigma^{\gamma^*\pi^0}_N+
\smallfrac{1}{6}\sigma^{\gamma^*\eta}_N +
\third\sigma^{\gamma^*\eta^\prime}_N-
\sigma^{\gamma^*\pi^+}_N\right)\cr &\simeq
{d\over dt}
\left(\smallfrac{1}{6}\sigma^{\gamma^*\eta}_N +
\third\sigma^{\gamma^*\eta^\prime}_N-
\half\sigma^{\gamma^*\pi}_N\right).\cr}}
The non--perturbative
contribution to the evolution factor for the Gottfried sum
\gottevol\  is thus
\eqn\evolnonpert
{\eqalign{\left[{\Delta}_1^+(t,t_0)\right]_{\rm nonpert}=
\exp &\bigg(
\left[\smallfrac{1}{6}\sigma^{\gamma^*\eta^\prime}_1(t)
+\third\sigma^{\gamma^*\eta^\prime}_1(t)
-\half\sigma^{\gamma^*\pi}_1(t)\right]\cr
&\qquad -\left[\smallfrac{1}{6}\sigma^{\gamma^*\eta^\prime}_1(t_0)
+\third\sigma^{\gamma^*\eta^\prime}_1(t_0)
-\half\sigma^{\gamma^*\pi}_1(t_0)\right]
\bigg).\cr}}

Whereas the perturbative evolution \evolpert\ was very small
the non--perturbative scale factor \evolnonpert\ is potentially large.  In
particular, the role played by the anomalous $U(N_f)$ symmetry
breaking in the pseudoscalar spectrum is now apparent: because of the large
$\eta$-$\pi$ and $\eta^\prime$-$\pi$ mass differences we expect the
corresponding emission cross sections to be significantly different.
In the $U(N_f)$-symmetric limit all cross sections would be the same
and the \rhs\ of eqn.\evolud , and thus the exponent in
eqn.\evolnonpert , would be identically equal to zero.
In reality, whereas for very large values of $Q^2$ the effects of the
pseudoscalar mesons' masses should be negligible and
the non--perturbative scale factor should flatten (as $1/Q^2$), for
scales of the order of the nucleon mass the effects of the mass
splitting will be so large that the $\eta$ and $\eta^\prime$
production cross sections will be negligible when
compared to that of the $\pi$. Thus because of the axial anomaly the diagonal
radiation process is disfavoured compared to the non--diagonal one,
as  half of it proceeds through $\eta$ and $\eta^\prime$ emission,
which is dynamically suppressed. So not only is the evolution purely
multiplicative, but since the cross--sections are all positive,
it necessarily results in the screening of the Gottfried
sum; eqn.\evolnonpert\ shows that $S_G$ is always
reduced as $Q^2$ is increased.

In order to make these observations quantitative, we must proceed to the
computation of the various radiation cross sections and the determination of
their first moment. We will do this in the next section.

\newsec{Non--perturbative Splitting Functions}

The computation of the non--perturbative splitting function reduces to
the computation of the cross section for the process $\gamma^* q\to \Pi q$,
where $\Pi$ is a pseudoscalar meson, which at leading order is given by the two
diagrams displayed in fig.~3. In order to perform this computation we must
introduce an effective quark--pseudoscalar coupling. Writing the quark
propagator as $S^{-1}(p)=(\pslash +\Sigma(p^2))^{-1}$, where
$\Sigma (p^2)$ is the quark self energy, the most general such
coupling is, by definition,
\eqn\bsdef
{\chi (k,p)\equiv S^{-1}(p+\half k)
\matele{0}{\psi(p-\half k)\bar{\psi}(p+\half k)}{\Pi(k)}S^{-1}(p-\half k),}
where $|\Pi(k)\rangle $ is the meson state, and the momenta are
notated as in fig.~2. The quark dynamics may thus be summarized by
writing an effective action
\eqn\seff
{S_{\rm eff}=\int\!d^4p\;\bar\psi (p)i\slash{D}\psi(p)+
\int\!d^4p\int\!d^4k\;\bar\psi(p+\half k)f\chi(k,p)U(k)\psi(p-\half k),}
where $U(k)\equiv\exp\big(i\gamma_5\Pi(k)/f_\Pi\big)$
is a nonlinear representation of the pseudoscalar field $\Pi(k)$;
expanding $U$ in powers of
$\Pi$ gives the quark self energy and then couplings of the
$q\bar{q}$ pair to increasing numbers of pseudoscalars $\Pi$.
In the chiral limit (i.e. ignoring quark masses)
the quark self--energy and the on--shell coupling to pions are thus related
by the chiral Ward--Takahashi identity
\eqn\severt
{\Sigma(p)=f\chi (0,p) ,}
$f$ being the meson decay constant
(see for example \ref\jj{R.~Jackiw and K.~Johnson,
\PR\vpy{D8}{2386}{1973}.}). Effective actions such as eqn.\seff\ have long been
used to discuss dynamical chiral symmetry breaking and to compute meson
form factors and decay amplitudes
\nref\chls{C.~H.~Llewellyn~Smith {\it Ann. Phys. (NY)}
\vpy{53}{521}{1969}.}\nref\pags{H.~Pagels,
\PR\vpy{D15}{2991}{1977}; \vpy{D19}{3080}{1979};
\vpy{D21}{2337}{1980}.}\nref\pagsto{H.~Pagels and S.~Stokar,
\PR\vpy{D20}{2947}{1979}.}\refs{\chls{--}\pagsto};
more recently attempts have been made to find a formal justification
for them within QCD (\ref\msb{R.~D.~Ball, {\it Int. Jour. Mod.
Phys.}\vpy{A5}{4391}{1990}.} and references therein) and to develop their
low energy phenomenological implications more systematically
\ref\hol{B.~Holdom, \PR\vpy{D45}{2535}{1992}\semi R.~D.~Ball and
G.~Ripka, in preparation.}.

The effective coupling \bsdef\ may be expanded in terms of a set
of four scalar vertex functions $\phi_i(k,p)$, $i=0,1,2,3$;
\eqn\vertdef
{\chi (k,p)=\phi_0(k,p)+\phi_1(k,p)\kslash+\phi_2(k,p)(k\cdot p)\pslash
+\phi_3(k,p)\half(\pslash\kslash -\kslash\pslash ).}
The vertex functions are all even functions of $k\cdot p$, and for
on--shell pseudoscalars we may expand them as
\eqn\vertexpand
{\phi_i(k,p)=\sum_{n=0}^{\infty}\phi_i^n(p^2)(k\cdot p)^{2n}.}
Now, in principle, all of the on--shell vertex functions $\phi_i^n(p)$
may be computed by solving the homogeneous bound state equation,
separated out into a hierarchy of equations corresponding to
increasing powers of $k$. In practice of course this is not possible
since the quark--antiquark scattering kernel is unknown except
at large Euclidean momentum transfer. The
asymptotic behaviour for large Euclidean $p^2$ is given by\ref
\btbse{R.~D.~Ball and J.~Tigg, unpublished.}
\eqn\vertasymp
{\eqalign{\phi_0^n(p^2)&\p2inf\frac{(\log p^2)^{\delta_0^n}}
{(p^2)^{n+1}},\cr
\phi_1^n(p^2)\p2inf p^2\phi_2^n(p^2)&\p2inf
\frac{(\log p^2)^{\delta_1^n}}{(p^2)^{n+1}},
\qquad\phi_3^n(p^2)\p2inf \frac{(\log p^2)^{\delta_3^n}}{(p^2)^{n+2}},\cr}}
where the numbers $\delta_i^n$ are related to the anomalous dimensions
of various local operators, and decrease monotonically in $n$.
Asymptotically, the dominant vertices are
thus $\phi_i(p^2)\equiv\phi_i^0(p^2)$ for $i=0,1,2$. Furthermore, it
can be seen from examination of the bound state equations that all
the $\phi_i^n(p^2)$ remain finite as $p^2\rightarrow 0$.

The normalization of the vertex functions is fixed by the
chiral Ward--Takahashi identity \severt\ in terms  of the
self--energy and $f_\pi$:
\eqn\WTI
{\phi^{\pi}_0(p^2)=\Sigma (p^2)/f_{\pi}.}
The scale of $\Sigma (p^2)$, in turn, is in principle given by
solution of the (nonlinear) Schwinger--Dyson equation in terms of
$\Lambda _{QCD}$. Infrared uncertainties make such a determination
unrealistic, however\btbse, and it is more practical to use the
Pagels--Stokar condition\pagsto\ which normalizes $\Sigma (p^2)$ to the pion
decay constant $f_\pi=93$~MeV:
\eqn\ffnorm
{f_{\pi}^2={N_c\over 4\pi^2}\int_0^{\infty}\!dp^2\,
p^2 \Sigma (p^2) \frac{\Sigma (p^2)-\half p^2
\frac{d \Sigma (p^2)}{dp^2}}{\left(p^2+\Sigma (p^2)^2\right)^2}.}
Then, using eqn.\WTI\ in eqn.\ffnorm\ fixes the strength of the meson--quark
coupling, or (more precisely) the normalization of the vertex function
in terms of the scale of the quark self--energy.
A consistency check on the condition \ffnorm\ can be obtained by
considering the asymptotic behaviour of the pion electromagnetic
form factor (also derived in ref.\pagsto):
\eqn\ffemff
{Q^2F_{\pi}(Q^2)\Q2inf \frac{N_c\ln 2}{4\pi^2f_{\pi}^2}\int_0^{\infty}
\!dp^2\,p^2\frac{\Sigma (p^2)^2}{\left(p^2+\Sigma (p^2)^2\right)}.}
In principle all the other vertex functions may now be found using
the homogeneous Bethe--Salpeter equations without the need for
further normalization conditions,
though in practice this is again not really viable since they are
too sensitive to the unknown infrared dynamics.

It is now straightforward to compute the spin--averaged
cross section corresponding to the two diagrams of fig.~3.
Assuming that all the quark partons are massless, writing $M$ for the
mass of the emitted meson, and $-Q^2$ for the virtuality of the
incident photon, the usual Mandelstam invariants $\hat s$, $\hat t$,
$\hat u$, are given, in terms of the four external particles'
four--momenta indicated in fig.~3, by
\eqn\mandl
{\eqalign{\hat s=&\left(p_i+q\right)^2=\left(p_f+k\right)^2
=-Q^2+2p_i\cdot q=M^2+2p_f\cdot k,\cr
\hat t=&\left(p_i-k\right)^2=\left(p_f-q\right)^2
=-Q^2-2p_f\cdot q=M^2-2p_i\cdot k,\cr
\hat u=&\left(p_i-p_f\right)^2=\left(q-k\right)^2
=-2p_i\cdot p_f= M^2-Q^2-\hat s-\hat t.\cr}}
The physical kinematical region is $\hat s\geq M^2$,
$\hat t\leq 0$, $\hat u\leq 0$.
Introducing the appropriate kinematical factors\foot{The cross
section is dimensionless after the extraction of a
scale factor which we fix in such a way that in the limit in which the meson is
point--like the canonically normalized splitting function \alt\
for emission of scalar particles is reproduced.} then,
after a lengthy computation, the
cross section reduces to the relatively simple form
\eqn\xsec
{\eqalign{\sigma^{\gamma^*\Pi}_{qq}(\hat s,\hat t)=-{1\over 16 \pi^2}
\Biggl[&\varphi_{\hat s}^2 \left({\hat t\over \hat s}+{Q^2M^2
\over {\hat s}^2}\right)+
\varphi_{\hat t}^2 \left({\hat s\over \hat t}+{Q^2M^2\over {\hat t}^2}
\right)+\cr
&\qquad2 \varphi_{\hat s} \varphi_{\hat t} \left(1-{M^2\over \hat s}
\right)\left(
1-{M^2\over \hat t}\right)+\left(\tilde\varphi_{\hat s}- \tilde
\varphi_{\hat t}\right)^2 \hat u\Biggr],
\cr}}
where
\eqn\newvertdef
{\eqalign{\varphi_{\hat s}&\equiv
\sum_{n=0}^\infty
\left(\half \hat s\right)^{2n}\left[\phi_0^n\big(\quarter (2\hat s-M^2)\big)
 +\half (\hat s-M^2)\phi_3^n\big(\quarter(2\hat s-M^2)\big)\right],\cr
\tilde\varphi_{\hat s}&\equiv
\sum_{n=0}^\infty
\left(\half \hat s\right)^{2n}\left[\phi_1^n\big(\quarter (2\hat s-M^2)\big)
 +\quarter \hat s\phi_2^n\big(\quarter(2\hat s-M^2)\big)\right],\cr
}}
and similarly for $\varphi_{\hat t}$ and $\tilde\varphi_{\hat t}$.

The cross section in terms of scaling variables is obtained by expressing
the Mandelstam invariants in terms of $x$, $Q^2$, and the center--of--mass
scattering angle $\theta$.  In the limit of vanishing quark mass (but
retaining of course the meson's mass) we have
\eqn\kin
{\eqalign{&\hat s={1-x\over x}Q^2,\cr
&\hat t= -\frac{(1-x)Q^2-xM^2}{2x(1-x)}(1-\cos\theta)-\frac{xM^2}{1-x},\cr
&\hat u=-\frac{(1-x)Q^2-xM^2}{2x(1-x)}(1+\cos\theta).\cr}}
The physical region corresponds thus to $0\le x\le x_{\rm max}$, with
$x_{\rm max}$ given by eqn.\xmeq.
It is now easy to check that when $x=x_{\rm max}$,  $\hat s=M^2$,
$\hat t=-Q^2$ and $\hat u=0$, so $\sigma^{\gamma^*\Pi}_{qq}$ vanishes
identically, as we claimed following eqn.\npad.
Integrating $\sigma^{\gamma^*\Pi}_{qq}(\hat s,\hat t)$ over $\cos\theta$
from $-1$ to $1$ then gives the
cross--section $\sigma^{\gamma^* \Pi}_{qq}(x;\ln Q^2)$ which enters the
expression \npad\ for the anomalous dimensions, and thus \evolnonpert\
for the evolution of the Gottfried sum.

Before we can do this, however, we must decide on a particular
form for the vertex functions. From the asymptotic behaviour
\vertasymp , and the definitions \newvertdef\ we may infer that
\eqn\newvertasymp
{\varphi_{\hat s}\twiddles{\hat s}\frac{\big(\log \hat s\big)
^{\delta_0^0}}{\hat s},
\qquad\tilde\varphi_{\hat s}\twiddles{\hat s}
\frac{\big(\log \hat s\big)^{\delta_1^0}}{\hat s};}
we also know that both remain finite at small $\hat s$.
We use these facts to justify
the simple ans\"atze\foot{Even though the asymptotic
behaviour \vertasymp\ may only be proven rigorously for spacelike
$p^2>0$ we assume that it may be analytically continued to timelike
$p^2<0$; we will need vertex functions in both spacelike and timelike
regions, since although $\hat s>\half M^2$, $\hat t<\half M^2$ (see \kin ).}
\hbox{$\varphi_{\hat s}=\varphi\big(\quarter (2\hat s-M^2)\big)$},
\hbox{$\varphi_{\hat t}=\varphi\big(\quarter (M^2-2\hat t)\big)$} with
\eqn\ffp
{\varphi^{\pi}(p^2)=\frac{m_d}{f_\pi}\frac{\Lambda^2+m_d^2}{\Lambda^2+p^2},
\qquad\tilde\varphi^{\pi}(p^2)=
\frac{g_\pi}{f_{\pi}}\frac{\tilde\Lambda^2+m_d^2}{\tilde\Lambda^2+p^2}.}
Similar forms will be assumed for the $\eta$ and $\eta '$ vertex
functions. We also make the simplifying assumption (strictly justifiable only
asymptotically) that $\Sigma (p^2)=f_{\pi}\varphi ^{\pi}(p^2)$.

The parameters $\Lambda$ and $\tilde\Lambda$ give the scale of
transition  of the vertex functions from the hard (pointlike) to the
soft (extended) region; they may be thought of as parameterizing the
``size'' of the bound state. The  normalization condition eqn.\ffnorm\
fixes the strength of the pion--quark coupling by fixing  $m_d$ as a
function of $\Lambda$. Because $m_d$ may be identified with the dynamical
quark mass (actually the ``pole''
mass, since $\Sigma (m_d^2)=m_d$) $\Lambda$ could then be fixed by requiring
this to be equal to the constituent quark mass. However, given the large
uncertainty on this quantity, this only gives us a range of acceptable values
of $\Lambda$, rather than fixing it completely. There is no comparable
way of fixing $g_{\pi}$, so we must treat $g_{\pi}$ and $\tilde\Lambda$
as free parameters; the dependence of the cross--section\xsec\ on
these is however rather weak except in the large $Q^2$ region, as we
will see in detail below.

Eqn.\xsec\ shows that the cross section is dominated by the region where either
the $t$-channel or  the $s$-channel\foot{The fact that the $s$-channel diagrams
can also be resummed and factorized just as the $t$-channel ladder diagrams is
discussed for a theory of scalar gluons in ref.\CHLStwo.}
intermediate particle goes close to the mass
shell (which it can never reach because of the meson mass) --- the
region where either $\hat t$ or $\hat s$ is small. Physically, this
corresponds to the emitted meson and the quark being (almost)
collinear in the respective diagrams fig.~3a and fig.~3b,
consistent with the probabilistic interpretation. In this region the form
factors eqn.\newvertasymp\ are essentially pointlike if the mesons are light.
Now, eqn.\kin\ shows that in the $t$-channel this
occurs when $\theta= 0$ and  $x$ is small, whereas in the $s$-channel
it occurs when $x$ is close to $x_{\rm max}$ (as given by
eqn.\xmeq). Both conditions are
valid for all $Q^2$, even though the phase-space region (i.e., the $x$-range)
where $\hat s$ and $\hat t$ are close to their respective
minimal values gets smaller as $Q^2$ increases.

This implies that for moderate $Q^2$ the cross section
eqn.\xsec\  is not only significantly increasing with $Q^2$, but also it is
dominated by the collinear region where the form factors are close to pointlike
and the dominance of ladder diagrams is valid. As $Q^2$ increases the
cross-section increases more slowly. Even though the bulk of the
cross section still comes from the collinear region, its growth
is effectively cut by the extended tail of the form factor, thus
leading  to
anomalous dimensions which have a ``higher--twist'' fall--off as inverse powers
of $Q^2$, until eventually the nonperturbative cross-section is negligible as
compared to the perturbative one. The cross-section \xsec\  with
form factors behaving asymptotically as \newvertasymp\ thus
provides a smooth interpolation between the region where
strong nonperturbative evolution takes place, and the asymptotic perturbative
region\foot{
If instead of \ffp\ we were to take the cruder forms
$\varphi (p^2)=\theta (\Lambda^2-p^2)$, $\tilde\varphi (p^2)=0$ we would,
after a chiral rotation, recover the chiral quark model\gm\ as used in
\ehq. This is not an acceptable approximation
precisely because it is the $1/p^2$ fall--off of
$\varphi (p^2)$, as embodied in \ffp, which gives the higher
twist evolution at large $Q^2$.}.

So far we have taken all the quarks to be massless; the resulting cross
section \xsec\ then diverges at small $\hat s$ or small $\hat t$, and,
as we discuss in the next section,
this means that the evolution due to Goldstone bosons diverges
in the chiral limit $M\rightarrow 0$. This is unacceptable, but easily cured.
Since we have effectively included radiative corrections to the
two--loop perturbative evolution diagram of fig.~2 which turn the emitted
(and unobserved) $q\bar q$ pair into a bound state (a meson)
we should include also radiative corrections on the propagating
quark\foot{Though not of course on the external ones.}.
Indeed, this is required for consistency:
since we are treating the nonsinglet pseudoscalars as Goldstone
bosons, which become massless in the chiral limit, chiral symmetry
must have been broken dynamically, and the propagating
quarks must have a dynamically generated self energy $\Sigma(p^2)$
(as may be seen directly from eqn.\WTI) of the order of several
hundreds of MeV.

The quark self energy corrections can be absorbed
into the vertex functions by the simple replacement
\eqn\newBSamp
{\chi(p_f+\half k,k)\rightarrow \frac{\hat s-(\pslash_i +\qslash )
\Sigma(\hat s)}{\hat s+\Sigma^2(\hat s)}\chi(p_f+\half k,k)}
in the $s$--channel, and similarly for the $t$--channel.
By rewriting
these new amplitudes in the form \vertdef, it is not difficult to see
that the cross--section retains its simple form \xsec, but now with
\eqn\sigvertdef
{\eqalign{\varphi_{\hat s}&\equiv
\frac{\hat s}{\hat s+\Sigma^2(\hat s)}\sum_{n=0}^\infty
\left(\half \hat s\right)^{2n}\left[\phi_0^n +\half (\hat s-M^2)\phi_3^n
 +\big(\phi_1^n+\quarter \hat s\phi_2^n\big)\Sigma (\hat s)\right],\cr
\tilde\varphi_{\hat s}&\equiv
\frac{1}{\hat s+\Sigma^2(\hat s)}\sum_{n=0}^\infty
\left(\half \hat s\right)^{2n}\left[\hat s\big(\phi_1^n +\quarter \hat
s\phi_2^n\big)
+\big(\phi_0^n +\half (\hat s-M^2)\phi_3^n\big)\Sigma (\hat s)\right],\cr
}}
where again all of the $\phi_i^n$ are to be evaluated at $\quarter
(2\hat s-M^2)$, and $\varphi_{\hat t}$, $\tilde\varphi_{\hat t}$ are
obtained by replacing $\hat s$ with $\hat t$. Although the asymptotic
behaviour \newvertasymp\ is
unchanged, in the infrared we now find that, since $\Sigma (\hat s)$
is finite for small $\hat s$, $\varphi_{\hat s}$ falls to zero linearly
with $\hat s$.
This is, as we show explicitly below, just sufficient to cure the
infrared divergence by suppressing the emission of Goldstone bosons in the
chiral limit. To incorporate this behaviour, we take the new ans\"atze
\eqn\xmass
{\varphi_{\hat s}=\frac{\hat s}{\hat s+m_d^2}\varphi\big(
\quarter (2\hat s-M^2)\big),\qquad
\varphi_{\hat t}=\frac{\hat t}{\hat t-m_d^2}\varphi\big(\quarter (M^2-2\hat
t)\big),}
with $\varphi (p^2)$ given by \ffp , and
$\tilde\varphi_{\hat s}$, $\tilde\varphi_{\hat t}$ as before.

On the other hand, consistency with the partonic interpretation requires that
quarks are treated as massless or quasi--massless. Hence, the present approach
will only be justified if the pion mass is sufficiently small that the
effects of chiral symmetry breaking in the pseudoscalar meson sector are felt,
but also sufficiently large that the inclusion of the quark self
energy in the cross-section is not
quantitatively important. This will have to be checked explicitly below.

\newsec{Computing the Evolution of the Gottfried Sum}

The $Q^2$ dependence of the Gottfried sum is obtained by integrating
the meson emission cross--section, eqn.\xsec,
over $\cos\theta$, taking the first moment, and then using the result
to evaluate the evolution factor \evolnonpert.
The full analytic expression for the integrated cross--section
with the form factors \ffp\ is given
in the appendix. Even though the general analytic form  eqn.(A.1) of
the integrated cross section is rather cumbersome, we may understand
the main qualitative
features of the first moment by considering various limits of it.

In the large-$Q$ limit the first moment of the
cross section for any finite $M$ behaves as
\eqn\largeq
{\sigma_1^{\gamma^*\Pi}(Q^2){\mathop\sim\limits_{Q^2\to \infty}}
\frac{m_d^2(m_d^2+\Lambda^2)^2}{\pi^2 f_\pi^2}
\left[k_0+\frac{k_1}{Q^2}+O\left(\frac{1}{Q^4}\right)\right],}
where $k_0$ and $k_1$ are
dimensionless functions of $M$, $m_d$, $g_\pi$, $\Lambda$
and $\tilde \Lambda$.
We thus have power suppression of the anomalous dimension
at large $Q^2$, i.e., ``higher--twist'' behaviour.
This is of course a direct consequence of the presence of the vertex functions
$\phi_i(k,p)$ eqn.\vertdef\ in the quark--meson coupling \seff, which forces
a deviation from the pointlike behaviour of the emission vertex at
large $Q^2$. The $1/Q^2$ falloff is a direct consequence of the soft
asymptotic behaviour \vertasymp ; if we had included the logarithms in
the ans\"atze \ffp , we would also have found logarithmic corrections to
the higher--twist behaviour.
Notice that the anomalous dimension of the Gottfried sum vanishes more rapidly
at large $Q$ than each of the anomalous dimensions for $\eta$ and $\pi$
emission does, because  the former is proportional to the difference of the
latter two.

Furthermore, it is easy to see that the behaviour of the first moment in the
large and small meson mass limits at fixed $Q^2$ is widely different.
Indeed, as the meson mass $M$ tends to zero (the chiral limit) the splitting
function reduces to that for the emission of a pointlike pseudoscalar
particle, namely (cf eqn.(A.5))
\eqn\smallm
{\sigma_1^{\gamma^*\Pi}( Q^2){\mathop\sim\limits_{M^2\to 0}}
\frac{1}{32\pi^2}\frac{m_d^2(m_d^2+\Lambda^2)^2}{\Lambda^4 f_\pi^2}
\log\left({Q^2\over M^2}\right).}
This would lead to singular behaviour in the chiral limit due to the usual
collinear divergence when the quark propagator in the diagrams of fig.~3
goes on--shell. When the meson mass tends to zero, however, we can no longer
neglect the quark self energy, which, due to  eqn.\xmass\ regulates
this divergence in the infrared. We discuss the chiral limit in
more detail below; here we merely remark that eqn.\smallm\ shows that when
the emitted meson is very light the anomalous dimension obtained from
eqn.\lightev is constant, and the scale dependence of all moments
is strong.

In the opposite limit of very large meson mass
the first moment of the cross section vanishes as
\eqn\largem{\eqalign{\sigma^{\gamma^*\Pi}_1(Q^2)
&{\mathop\sim\limits_{M\to\infty}}
\frac{2g_\pi^2(m_d^2+\tilde\Lambda^2)^2}{\pi^2 f_\pi^2}\frac{Q^2}{M^4}
\left[1+O\bigg({1\over M^2}\bigg)\right]\cr
&\qquad\qquad+{m_d^2(m_d^2+\Lambda^2)^2\over f_\pi^2}\frac{Q^2}{M^6}
\Big(\ln \bigg({M^2\over Q^2}\bigg)+c_2\Big)
\left[c_1+O\bigg({1\over M^2}\bigg)\right],\cr}}
where $c_1$ and $c_2$ are dimensionless functions of
$Q^2$, $m_d$, $g_\pi$, $\Lambda$ and $\tilde \Lambda$.
Heavy mesons thus make only a very small contribution to the
evolution, as expected. The fall--off of
moments of the cross--section at large $M$ is so rapid because
although the cross--section itself only falls as $1/M^2$,
$x_{\rm max}\sim Q^2/M^2$. It is interesting to notice
that the leading contribution is proportional to $g_\pi^2$,
because the large $M$ expansion of the cross--section begins
at order $x$ when $g_\pi=0$ (see eqn.(A.6)).

Summarizing, an analysis of various limits of  the anomalous dimension
computed from the cross section for meson emission confirms the
naive expectations discussed in the end of section 2: due to the soft vertex
functions, at large $Q^2$ (compared to all other scales, including
$M$) the anomalous dimension always reduces to
higher--twist behaviour; for small meson mass (compared to all the
other scales, and in particular $Q^2$) the first moment evolves
strongly, while for large mass the anomalous
dimension is very small, implying that the nonsinglet anomalous dimensions
eqn.\evolud, and in particular the combination which governs the
evolution of $S_G$, are rather large.

We proceed now to a detailed computation of the evolution of the Gottfried
sum. In order to do this, we must still fix the free parameters in the
cross section eqn.\xsec. As we have seen in the previous section, $m_d$
may be determined as a function of $\Lambda$ by the normalization
condition eqn.\ffnorm, or the expression \ffemff, with the left hand
side determined experimentally to be $0.38\pm0.05\;\gev^2$\ref
\emff{C.~J.~Bebek et al., \PR\vpy{D13}{25}{1976}; \vpy{D17}{1693}{1978}.}.
The resulting dependence of $m_d$ on $\Lambda$ is shown in fig.~4, and turns
out to be rather weak over a reasonable range of $\Lambda$;
moreover, the two independent determinations eqn.\ffnorm\ and \ffemff\ are
compatible within the given uncertainties. In what follows we use for
definiteness the value of $m_d$ determined as a function of $\Lambda$
by eqn.\ffnorm, which (see fig.~4) provides us with a lower bound on the
anomalous dimension and therefore a conservative estimate of the evolution of
the Gottfried sum. $\Lambda$ is then treated as a free parameter; if $m_d$
is to be of the order of the constituent quark mass
($m_d\approx\half M_\rho\approx\third M_p$), $\Lambda$ must be
roughly in the range $0.4 \lsim \Lambda \lsim 0.8\;\gev$.

We are thus left with the parameter $g_{\pi}$, which
gives the relative importance of the derivative
coupling to the pseudoscalar coupling of the meson to the quark,
eqn.\seff. The sign of $g_\pi$ is irrelevant, as is apparent from eqn.\xsec.
Solutions of truncated Bethe-Salpeter equations seem to
indicate that $\phi_1$ and $\phi_2$ are rather small\btbse, but fall
off on the same scale as $\phi_0$;
according to \newvertdef\ and \ffp\ this suggests $g_\pi < 1$ and
$\tilde\Lambda\simeq\Lambda$. However
if quark self energy corrections are included, then
$\phi_0$ also contributes to $\tilde\varphi$ (\sigvertdef), suggesting that
$g_{\pi}\sim 1$\foot{Although these contributions are either soft,
or suppressed by $\bar{m}/m_d$, where $\bar{m}$ is the light current
quark mass.}. We thus consider $0\leq g_\pi\leq 1$ to be a realistic
range of values, and take $\tilde\Lambda =\Lambda$.
This choice is actually immaterial for small values of $Q^2$, up to
$Q^2$ of order of a few GeV$^2$, since the contribution of the terms
proportional to $g_{\pi}$ is negligible in this region.
This is due to the fact that
terms proportional to $g_{\pi}$ yield a negligible contribution
to the cross section in most of phase space, and in particular vanish
in the region where the form factors are pointlike (and, more generally,
whenever their arguments are equal). However they have a slower
fall--off at large $Q^2$ than the rest of the cross--section (because
the piece of $K_1$ (eqn.(A.4)) which is proportional to $g_\pi^2$ grows
as $\ln Q^2$) and thus dominate it asymptotically.

We have thus computed numerically
the first moment of the integrated cross--section eqn.(A.1)
for $\pi^0$, $\pi^\pm$, $\eta$ and $\eta^\prime$ emission, and the derivatives
of these first moments with respect to $\ln Q^2$,
which give the anomalous dimension for nonsinglet evolution
according to eqn.\npad. The results are displayed
in fig.~5 (cross section) and fig.~6 (anomalous dimension),
where the dependence on the meson mass and on the
parameter $\Lambda$ (with $g_\pi=1$) are also shown.
It is apparent that the contribution to the anomalous dimension
from $\pi$ emission is indeed significantly larger than that from
$\eta$ emission, thereby leading to substantial evolution of the first
moment for values of $Q^2$ between $0.1$ and $100$ GeV$^2$.
At small $Q^2$ the cross--sections evolve very slowly; furthermore,
their derivatives tend to a common value, thereby leading to vanishing
of the anomalous dimension and flattening of the nonsinglet evolution.
For large $Q^2$ the cross section flattens; the precise value where this occurs
is controlled by the parameter $g_\pi$ and may vary from $Q^2 \sim 10$ GeV$^2$
for $g_\pi=0$ to $Q^2 \sim 10^3$ GeV$^2$ for $g_\pi=1$.

The relative importance of various contributions to the cross section in
different $Q^2$ regions, as well as the dependence
on $g_\pi$ are displayed in fig.~7. It appears that $t$-channel pseudoscalar
emission (fig.~3b) provides the leading contribution in the very small
$Q^2 \sim 0.01$---$0.1\gev^2$ region; $s$-channel emission
(fig.~3a) provides the dominant contribution in the small to intermediate
$Q^2 \sim 0.1$---$1\gev^2$ region; and the derivative coupling terms,
proportional to $g_\pi$, control the large $Q^2$ tail.
Comparing fig.~7a ($\pi^0$ emission) with fig.~7b ($\eta^\prime$ emission)
demonstrates explicitly the relatively slower fall-off \largem\ of the
derivative coupling terms compared to the nonderivative ones with
increasing meson mass $M$ .

The $x$-dependence of the $s$ and $t$ channel contributions which,
upon integration, lead to the total cross section of fig.~7a is
further displayed in fig.~8. This shows
that indeed the nonperturbative evolution is dominated by the collinear region
where the leading logarithmic approximation holds and the form factors are
almost pointlike. It is interesting to observe that in the $s$-channel the
collinear region is dominant despite the fact that this is the
large-$x$ region, and the cross-section vanishes  at $x=x_{\rm max}$.

The flattening of the evolution at small $Q^2$ identifies naturally a
scale at which physical observables should be connected smoothly to the
quark--model values, which one expects to hold in the infrared regime.
It is pleasing to notice that the flattening occurs at
$Q^2\sim (200{\rm ~MeV})^2$, the natural ``QCD scale'' at which
confinement and chiral symmetry breaking are expected to occur. Moreover the
consistency condition resulting from \consb\ can indeed be satisfied in this
region (and in no other); when $\gamma_1^{qq}\simeq 0$,
$\pi_1\simeq 0$.
At this scale there is no dynamically generated meson component, hence the
sea is expected to be symmetric and $S_G$ take its quark model, valence value
$S_G={1\over 3}$. This result cannot be compared directly with
those obtained in effective models (such as Skyrme or chiral quark
soliton models) because in such models there is no scale dependence.

We can thus proceed to compute the scale dependence of the
Gottfried sum assuming that it takes the quark model value $S_G={1\over 3}$
at a reference scale $Q_0= 200$~MeV, where the evolution flattens, and then
using eqn.\gottevol. We take $\Lambda=\tilde\Lambda$ and we let this parameter
vary in the range discussed above, while we fix $m_d$ from fig.~4 (solid line),
i.e. from eqn.\ffnorm; we then vary $0\le g_\pi\le 1$.
Good agreement with the experimental data
is obtained for $\Lambda\approx550$ MeV, as shown by fig.~9, where both the
dependence on $\Lambda$ and $g_\pi$ are shown.

Fig.~9 also displays the flattening of the evolution at both small and large
$Q^2$. Because of the multiplicative character of the evolution the
precise small-$Q^2$ value $Q_0$ at which the condition $S_G(Q_0)={1\over3}$ is
enforced is immaterial. Indeed, we have checked explicitly that if
instead of taking $Q_0=200$ MeV, we fix
$Q_0$ by requiring that the anomalous dimension be smaller than a threshold
value (so that $Q_0$ may depend on the specific values of the parameters)
the results are essentially unchanged. The value at which the evolution
flattens at large $Q^2$ is controlled by $g_\pi$, and is generally
smaller than the value at which the cross section for the emission of each
particle separately flattens, as we discussed above.
The flattening occurs thus typically around $Q^2\sim 3$ GeV$^2$
in the extreme case of $g_\pi=0$, and $Q^2\sim 100$ GeV$^2$ if $g_\pi=1$.

In any case, we predict that there is
a significant scale dependence of $S_G$ in the region of $Q^2\sim 1$ GeV$^2$,
which could be experimentally detectable. Furthermore, unless
$g_\pi\ll 1$, it appears that the asymptotic value of $S_G$ is not
yet attained in the region where present-day data are taken;
rather, the asymptotic value can be as low as
$S_G(\infty)\approx 0.20$ if $g_\pi\sim 1$.

We can now address the issue of the self-consistency of the present approach.
Firstly, the very fact of finding a positive value
of the anomalous dimensions is of itself nontrivial; there is no
fundamental reason why the cross section for meson emission should always
increase with $Q^2$. This provides an a posteriori
consistency check of the probabilistic interpretation of the splitting
functions which we have assumed throughout. Furthermore the flattening
of the cross--sections at small $Q^2$ is another nontrivial
consistency check; without it we could not assume that the meson
component of the sea is generated entirely by $Q^2$ evolution, due to
the condition \consb. The flattening at large $Q^2$ is consistent with
the asymptotic recovery of perturbative behaviour.
A study of the $x$ and $Q^2$ dependence of various contributions to
the cross--section (figs 7 and 8) shows that indeed significant
$Q^2$ dependence is generated nonperturbatively
only when the $t$-channel (at small $Q^2$) and the $s$-channel
evolution (at intermediate $Q^2$) are dominated by the respective
quasi--collinear singularities; this is consistent with the probabilistic
interpretation of the evolution equations.

Finally, let us consider the chiral limit of our approach. In this
limit the cross section eqn.\xsec\ diverges in the infrared because
the collinear singularities are no longer regulated by the meson mass.
This would lead to the behaviour eqn.\smallm\ of the splitting
function, and thus to
an anomalous dimension which does not display higher--twist behaviour at large
$Q^2$, but rather coincides with that obtained in a theory with pseudoscalar
pointlike gluons. This is untenable both theoretically and phenomenologically,
and would seem to suggest that the chiral limit reveals an
inconsistency in our approach.
However, as discussed at the end of section 4, in order to examine
the chiral limit consistently it is necessary
to compute the meson emission cross sections and the evolution of the
Gottfried sum by using the form eqn.\xmass\ for the vertex functions,
which corresponds to including the self--energy corrections in the
internal quark propagator. The external (parton) quarks are kept
massless, so the Mandelstam invariants retain their form eqn.\kin.

In order to simplify the computation, we have approximated the
function $\varphi_{\hat s}$ eqn.\xmass with the linearly rising function
$\varphi_{\hat s}={m_d^3\over f_\pi \Lambda^2}{\hat s\over m^2}$
for small $\hat s$, and the asymptotic form eqn.\ffp\ for large
$\hat s$, joined by demanding continuity. An analogous approximation
is taken for $\varphi_{\hat t}$. The dependence of the cross--section
on the pseudoscalar meson mass with and without
the inclusion of the self energy correction are
displayed in fig.~10. It may be seen that the self energy correction
does indeed succeed in taming the infrared divergence, thus giving
sensible results in the chiral limit.

However, the physical value of
$M_\pi$ is sufficiently large (and thus in practice far from the
chiral limit) that the self energy correction makes little difference
to the actual evolution of the Gottfried sum.
This is shown in fig.~11, where the evolution of the Gottfried sum with and
without self-energy corrections are compared (with different values of
$g_\pi$). It is thus
apparent that the partonic approach is indeed consistent with the
smoothness of the chiral limit, since our results for the
non--perturbative evolution of the Gottfried sum are essentially
insensitive to the dynamical quark mass generation.
It is interesting to observe that the normalization of the cross section
does depend on the propagating quark mass, but its derivative does not.

We conclude that our approach provides a fully self--consistent, rather
stable, and almost parameter--free determination of the anomalous evolution of
the Gottfried sum, which allows us to understand the current experimental data
in a natural way, and suggests a further $Q^2$ dependence of the Gottfried
sum which may be observable as and when new data at different
values of $Q^2$ become available.

\newsec{Conclusion}

In this paper we have presented an extension of the Altarelli--Parisi evolution
equation to include bound state emission, and computed the relevant splitting
functions in the particular case of pseudoscalar emission, which gives the
only significant contributions in the nonsinglet
channel, due to the large breaking of $U(N_f)$ flavor symmetry in the
pseudoscalar meson spectrum, and the relatively low mass of the
pseudo--Goldstone bosons. Our computational method is
based on the probabilistic interpretation of the Altarelli--Parisi equations,
and leads to a determination of the non--perturbative scale dependence
of nonsinglet quark distributions due to the direct coupling of the
quarks to the pseudoscalar bound states.
We have then determined explicitly the relevant splitting function,
we have computed its first moment, and we have determined the scale dependence
of the first moment of the nonsinglet quark distribution, and hence that of the
Gottfried sum.

The overall qualitative features of our results for the splitting functions
and for the evolution of the first moment confirm a posteriori the consistency
of our approach: the anomalous dimensions
are positive definite (consistently with the probabilistic interpretation),
and  flatten both at large $Q^2$, where they reduce
to higher--twist behaviour, and at small $Q^2$, around the confinement scale.
The evolution of the Gottfried sum is computed in terms of a single free
parameter, which is however fixed independently to within a factor of two,
and turns out to be exactly such as needed to explain the recent experimental
results, in that it joins smoothly to the quark model value at the confinement
scale, and reproduces the observed, significantly smaller value at
$Q^2=4$ GeV$^2$. The evolution has a consistent chiral limit, and is stable
upon variation of the parameters.

We believe that the results presented here provide an appealing and consistent
resolution of the puzzle posed by the recent data on the Gottfried sum; the
value of the sum which is measured experimentally is significantly smaller
than the quark model expectation because non--perturbative effects lead to
significant evolution in the intermediate $Q^2$ region.
This evolution produces a flavor asymmetry in the quark sea which screens the
value of the valence asymmetry, hence the value of the Gottfried sum.
The ultimate dynamical reason for this screening is the breaking of the
$U(N_f)$ flavor symmetry to $SU(N_f)$ in the pseudoscalar meson spectrum due to
the chiral anomaly. In this sense, the evolution of the Gottfried sum is
anomalous, just as the evolution of the first moment of the polarized structure
function $g_1^p$ \spinrev\ is expected to be \protball .
Our results also provide a definite, experimentally testable
prediction for the scale dependence of the Gottfried sum, and in particular
suggest that the asymptotic value of the Gottfried sum may be significantly
smaller than the present experimental determination.

Beyond the immediate application to the Gottfried sum, our results provide
a rather more general technique for the study of the non--perturbative
aspects of deep--inelastic scattering.
In particular, it will be interesting to compute
the evolution of the full nonsinglet quark densities (not just their
first moments) which, at scales below a few GeV, should be rather
different from what expected on the basis of perturbative QCD alone.
Indeed, there are now experimental indications that
the ratio of structure functions $F_2^n/F_2^p$ evolves differently
(i.e. significantly more strongly) than expected perturbatively\ref
\exprat{P.~Amaudruz et al., CERN preprint CERN-PPE/91-167 (1991).}.
Also, the present approach could shed new light
on the old idea\ref\radpart
{G.~Parisi and R.~Petronzio, \PL\vpy{B62}{331}{1976}
\semi V.A.~Novikov, M.A.~Shifman, A.I.~Vainshtein and V.I.~Zakharov,
{\it Ann.~Phys.~}\vpy{105}{276}{1977}
\semi M.~Gl\"uck and E.~Reya, \NP\vpy{B130}{76}{1977}.}
that the sea distributions are generated radiatively from simple valence
distributions at low scales where a quark model valence picture of the
nucleon is presumably valid. Whereas all previous attempts to this were
hampered by the obvious limitations of perturbative evolution,
the present approach could provide the required nonperturbative
information. In particular, it is encouraging to note that in contrast
to the perturbative evolution the
flatness of the non--perturbative evolution at low scales will greatly
reduce the sensitivity of the final distributions to the
(a priori unknown) starting scale.

The possibility of determining in this way the small-$x$ behaviour of the
nonsinglet distributions seems especially intriguing.
In the singlet channel, however,
the bulk of the evolution will be  controlled by the usual perturbative
anomalous dimensions, since the nonperturbative anomalous evolution
we determine, albeit unusually large for the nonsinglet channel, is
always small compared to the singlet perturbative evolution.
Nonetheless the effects which we discuss here may still be important
in peculiar kinematical ranges, such as very small $x$ at intermediate $Q^2$.

\bigskip
\noindent{\bf Acknowledgements:} SF thanks M.~Anselmino, V.~Barone,
L.~Magnea,
N.~N.~Nikolaev and E.~Predazzi for discussions. RDB would like to
thank the Royal Society for financial support.
\vfill
\eject

\Appendix{}

The cross--section $\sigma^{\gamma^*\Pi}(x;Q^2)$, obtained by integrating
the cross--section eqn.\xsec\ over $\cos\theta$ , is given by
\eqn\resxsec{
\eqalign{\sigma_{qq}^{\gamma^*\Pi}(x;Q^2)&=
-{1\over \pi^2}{m_d^2(m_d^2+\Lambda^2)^2\over f_\pi^2}
\Biggl\{{2Q^2(1 - x)\over\left( M^2 + 4\Lambda^2\right)^2}
{\left(4Q^2 - (3M^2-4\Lambda^2)x\right)\over (Q^2-M^2x)
[2Q^2 - (M^2 - 4\Lambda^2)x]}\cr
&\qquad\qquad\qquad\qquad\qquad +  {x (M^2x-Q^2(1-x)) \over
    Q^2(1 - x)^2[2 Q^2(1-x) - (M^2- 4\Lambda^2)x]^2}\cr
&\qquad+{1\over Q^2\left(M^2 + 4\Lambda^2\right)^3[Q^2(1 - x) -  M^2 x]
\left[2Q^2(1-x) -\left( M^2 - 4\Lambda^2\right)x\right]} \cr
&\qquad\times\Bigg[4C_1(x,Q^2,M^2,\Lambda^2)\ln\left({(1 - x)
\left(Q^2 - M^2x\right)\over M^2x^2}\right)\cr
&\qquad\qquad\qquad-C_2(x,Q^2,M^2,\Lambda^2)
\ln\left({x(M^2(1+x) + 4\Lambda^2(1-x))\over
        (1 - x)(2Q^2 - (M^2 - 4\Lambda^2)x)}\right)\Bigg]\cr
&\qquad + {g_{\pi}^2 \over m_d^2}
{(m_d^2+\tilde\Lambda^2)^2\over(m_d^2+\Lambda^2)^2}
\Bigg[{x (3 M^2-8\tilde\Lambda^2)-2(M^2-4\tilde\Lambda^2)x^2
-(1-x)(5-4x)Q^2 \over (1 - x)[2 Q^2(1-x) - (M^2-
4\tilde\Lambda^2)x]^2}\cr
&\qquad\qquad\qquad\qquad\qquad\qquad -{1-x\over [M^2(1+x) +
4\tilde\Lambda^2 (1-x)]} \cr
&\qquad\qquad +{x(1-x)[3(M^2-4\tilde\Lambda^2)x-2(3-x)Q^2] \over
2[Q^2(1 - x) - M^2x][2Q^2(1-x) - (M^2 - 4\tilde\Lambda^2)x]}\cr
&\qquad\qquad\qquad\qquad\qquad\times
\ln\left({x(M^2(1+x) + 4\Lambda^2(1-x))\over
        (1 - x)(2Q^2 - (M^2 - 4\Lambda^2)x)}\right)
\Bigg]\Biggr\},\cr}}
in terms of the parameters defined in section 2 \kin,\ffp
and with
\eqn\cdefs
{\eqalign{
C_1&(x,Q^2,M^2,\Lambda^2)=-2M^2(M^2+4\Lambda^2 )^2 x^2[M^2x-Q^2(1-x)]\cr
&\qquad +\left[M^2  +  4\Lambda^2 + (3M^2-4\Lambda^2)x \right](1-x)Q^4
 \left[(M^2-4\Lambda^2)x -Q^2(1-x)\right],\cr
C_2&(x,Q^2,M^2,\Lambda^2)=2M^2(M^2-4\Lambda^2)(M^2+4\Lambda^2)^2 x^2
\left[M^2 x -Q^2(1-x)\right]\cr
&\qquad\qquad -\left[ M^2+4\Lambda^2 +(3 M^2-4\Lambda^2)x\right]Q^4
\left[M^2-4\Lambda^2 +2Q^2(1-x)^2\right].\cr}}
In the limit of large $Q^2$ eqn.\resxsec\ gives
\eqn\reslargeq
{\sigma_{qq}^{\gamma^*\Pi}(x;\ln Q^2)=-{1\over \pi^2}
{ m_d^2(m_d^2+\Lambda^2)^2\over f_\pi^2}\left[
K_0+K_1 {1\over Q^2}\right]+O\left({1\over Q^4}\right),}
where
\eqn\kdefs
{\eqalign{ K_0&=
{4(1 - x)\over(M^2 + 4\Lambda^2)^2}
-{2[M^2(1+3x) + 4\Lambda^2(1-x)]
     \over (M^2 + 4\Lambda^2)^3}\ln\left({M^2(1+x) + 4\Lambda^2(1-x)
\over2M^2x}\right)\cr
&\qquad - {g_{\pi}^2 \over m_d^2}
{(m_d^2+\tilde\Lambda^2)^2\over(m_d^2+\Lambda^2)^2}
{(1 - x)\over [M^2(1+x) + 4\tilde\Lambda^2(1-x)]}\cr
K_1&=  { 4 M^2
  \over  (M^2 + 4\Lambda^2)^2 }
+ {2M^2x[M^2(1+3x) + 4\Lambda^2(1-x)]\over (1-x)
(M^2+4\Lambda^2)^3}
\ln\left({M^2(1+x) + 4\Lambda^2(1-x)\over 2M^2x^2}\right)\cr
&\quad +{g_{\pi}^2 \over m_d^2}
{(m_d^2+\tilde\Lambda^2)^2\over(m_d^2+\Lambda^2)^2}\bigg[
{x(5 - 4x)\over 4(1 - x)^2} -{x(3 - x)\over 2(1-x)}
\ln\left({x\left[M^2(1+x)+4\Lambda^2(1-x)\right]\over
2 Q^2(1-x) }\right)\bigg].\cr
}}
In the limit of small meson mass $M$ eqn.\resxsec\ yields
\eqn\ressmallm
{\sigma_{qq}^{\gamma^*\Pi}(x;Q^2)=
{1\over 16 \pi^2} {m_d^2(m_d^2+\Lambda^2)^2 \over
f_\pi^2\Lambda^4} (1-x) \ln\left({Q^2\over M^2}\right)+O\left(1\right),}
while in the limit of large meson mass $M$ eqn.\resxsec\ reduces to
\eqn\reslargem
{\eqalign{\sigma_{qq}^{\gamma^*\Pi}(x;Q^2)=&-{1\over\pi^2}
{ m_d^2(m_d^2+\Lambda^2)^2\over f_\pi^2 M^2}\Bigg\{
{x\over Q^2 (1 - x)^2} +{2x\over Q^2}\ln
\left({x^2\over 1 - x^2}\right)\cr
& + {g_{\pi}^2 \over m_d^2}
{(m_d^2+\tilde\Lambda^2)^2\over (m_d^2+\Lambda^2)^2}\bigg[
{2 + 3 x(1 - x)\over (1-x^2)}-{3 (1-x) \over 2}
\ln\left({1 - x\over1 + x}\right)\bigg]\Bigg\}+O\left({1
\over M^4}\right). \cr}}

The  expressions \reslargeq\ and \reslargem, however,
are not very useful in practice since, for large $Q^2$,
$x_{\rm max}=1-M^2/Q^2+O(M^4/Q^4)$ and, for large $M$,
$x_{\rm max}=Q^2/M^2+O(Q^4/M^4)$. To obtain the behaviour of the first
moment of \resxsec\ thus it is not enough to take moments of these equations;
rather one must expand consistently in both $1/Q^2$
and $x$, or, respectively, $1/M^2$ and $x$,
which leads to rather cumbersome expressions which will not be given here.
\vfill
\eject
\listrefs
\vfill
\eject
\centerline{\bf Figure Captions}
\bigskip
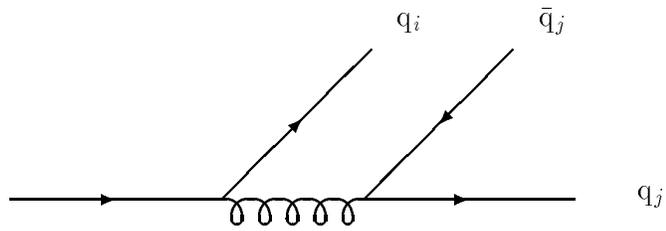
\item{[Fig.~1]} The two--loop diagram which generates nonsinglet evolution of
structure functions. When $i=j$ the final state must be antisymmetrized.
\medskip
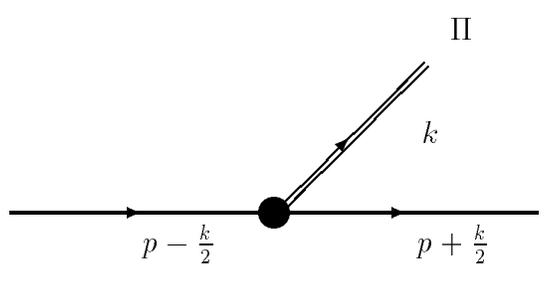
\item{[Fig.~2]} The diagram responsible for flavor symmetry breaking
non-perturbative evolution.
\medskip
\item{[Fig.~3]} Deep--inelastic scattering off a quark which radiates a
bound state $\Pi$: a)~$t$-channel diagram; b)~$s$-channel diagram.
\medskip
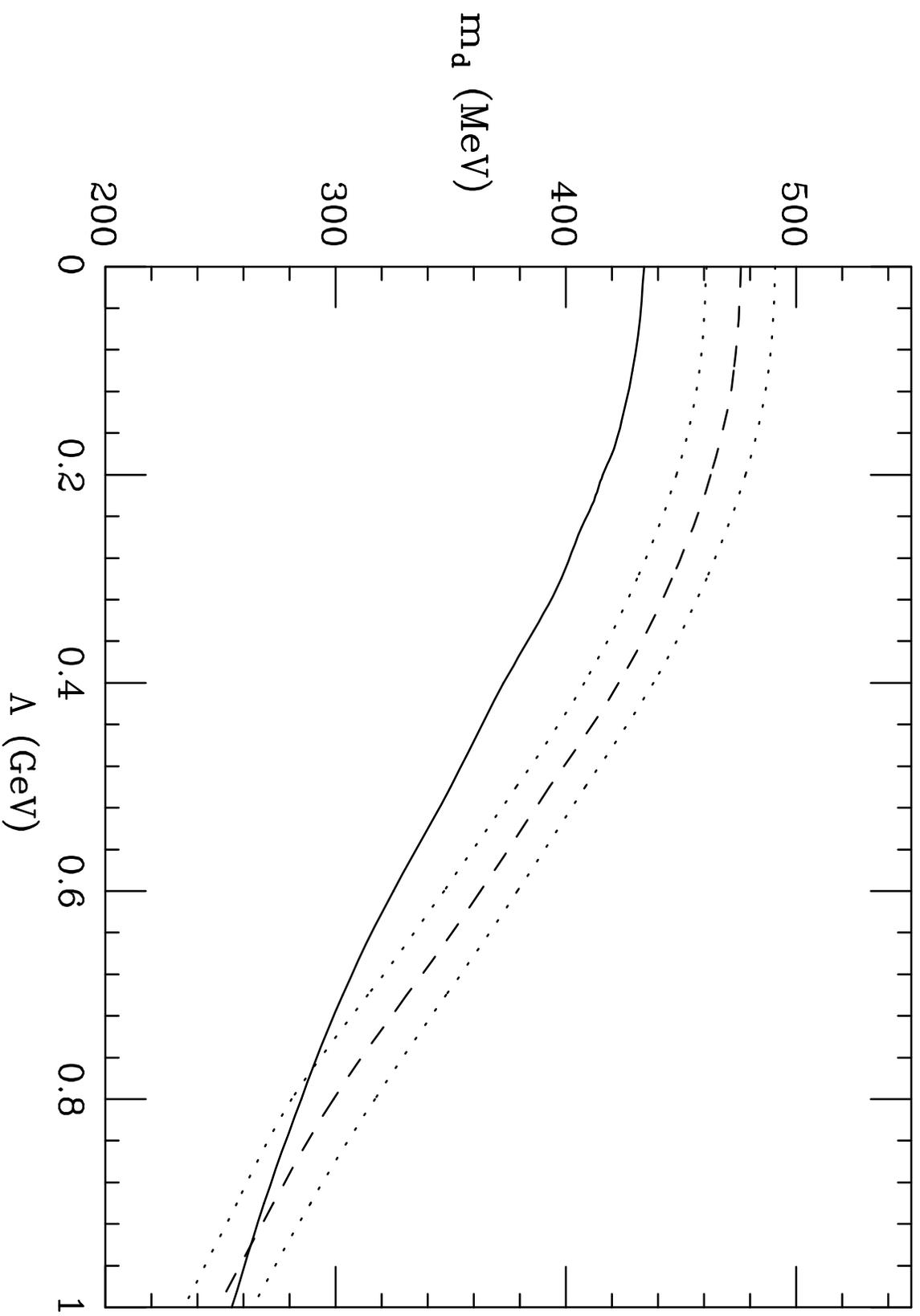
\item{[Fig.~4]} The constituent mass $m_d$ as a function of $\Lambda$,
as given by eqn.s \ffnorm\ (full line) and \ffemff\ (dashed line; the dotted
lines correspond to the experimental uncertainty).
\medskip
\item{[Fig.~5]} First moment $\sigma_1^{\gamma^*\Pi}$ of the cross-section
eqn.\xsec\ computed with $\tilde\Lambda=\Lambda$, $g_\pi=1$
and $m_d$ fixed from the full line of fig.~4: a) $\Pi=\pi^0$;
b) $\Pi=\eta^\prime$; c) $\Lambda=600$ MeV.
\medskip
\item{[Fig.~6]} Anomalous dimensions
$\gamma_1^{qq}={d\over dt}\sigma^{\gamma^*\Pi}_1$ computed from the
cross sections displayed in fig.~5.
\medskip
\item{[Fig.7]} Contributions of various terms in eqn.\xsec\ to the
cross section displayed in fig.~5a with $\Lambda=600$ MeV:
full line, cross section
with $g_\pi=0$; dot-dash line, cross section with $g_\pi={1\over 2}$;
dotted line, $s$-channel contribution (obtained setting $\phi_{\hat t}=g_\pi=0$
in eqn.\xsec); dashed  line, $t$-channel contribution (obtained setting
$\phi_{\hat s}=g_\pi=0$ in eqn.\xsec).
\medskip
\item{[Fig.~8]} Dependence on $x$ of the $s$-channel
and $t$-channel contributions whose integral with respect to $x$
is shown in fig.~7a, with several values of $Q^2$:
a) $t$-channel contributions (dotted lines of fig.~7a);
b) $s$-channel contributions (dashed lines of fig.~7a).
\medskip
\item{[Fig.~9]} Scale dependence of the Gottfried sum, computed with $\tilde
\Lambda=\Lambda$ and $m_d$ fixed from the solid curve in fig.~4: a)
$g_\pi=1$;
b) solid curves, $g_\pi=1$; dotted curves, $g_\pi={1\over2}$;
dashed curves, $g_\pi=0$. The experimental value eqn.\esum\
(uncorrected for shadowing) is also displayed.
\medskip
\item{[Fig.~10]} The approach to the chiral limit of the first moment
of the integrated cross section: a) Without
self-energy corrections; b) with self-energy corrections. The values of the
parameters are the same as in fig.~5c.
\medskip
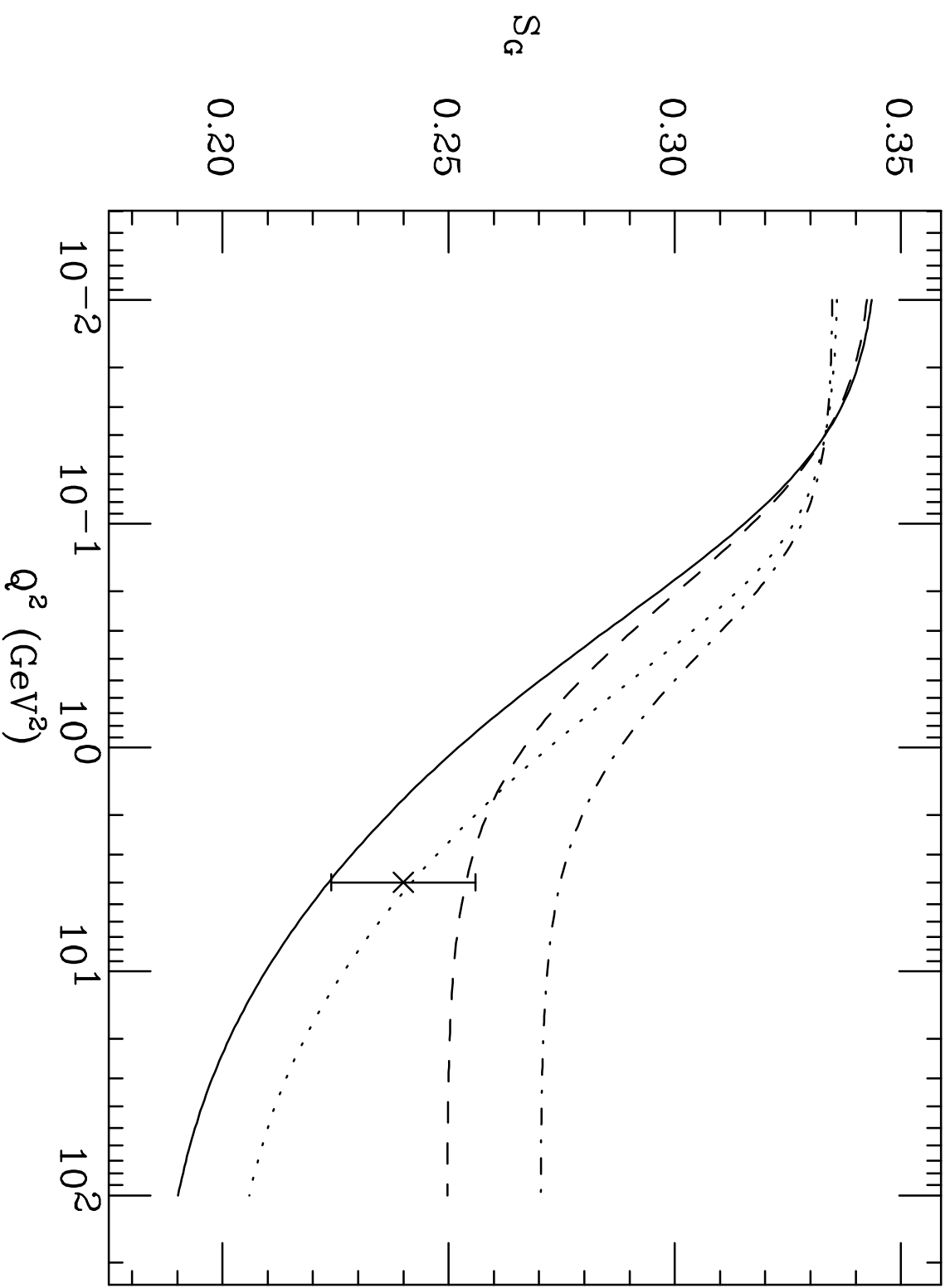
\item{[Fig.~11]} Detail of the scale dependence of the Gottfried sum:
solid curve, as the solid curve in fig.~9b ($g_\pi=1$); dotted curve,
same with
dynamical mass correction; dashed curve, as the dashed curve in
fig.~9b ($g_\pi=0$); dash-dot curve, same with the
dynamical mass corrections. The values of the parameters are the same as in the
curves of fig.~8 but with $\Lambda=500$ MeV.

\medskip
\vfill
\eject
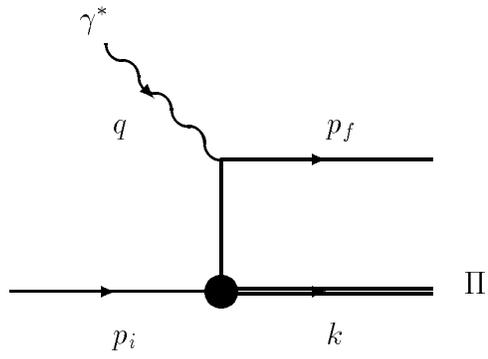
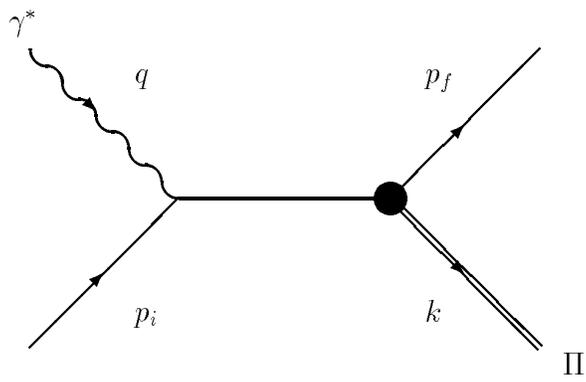
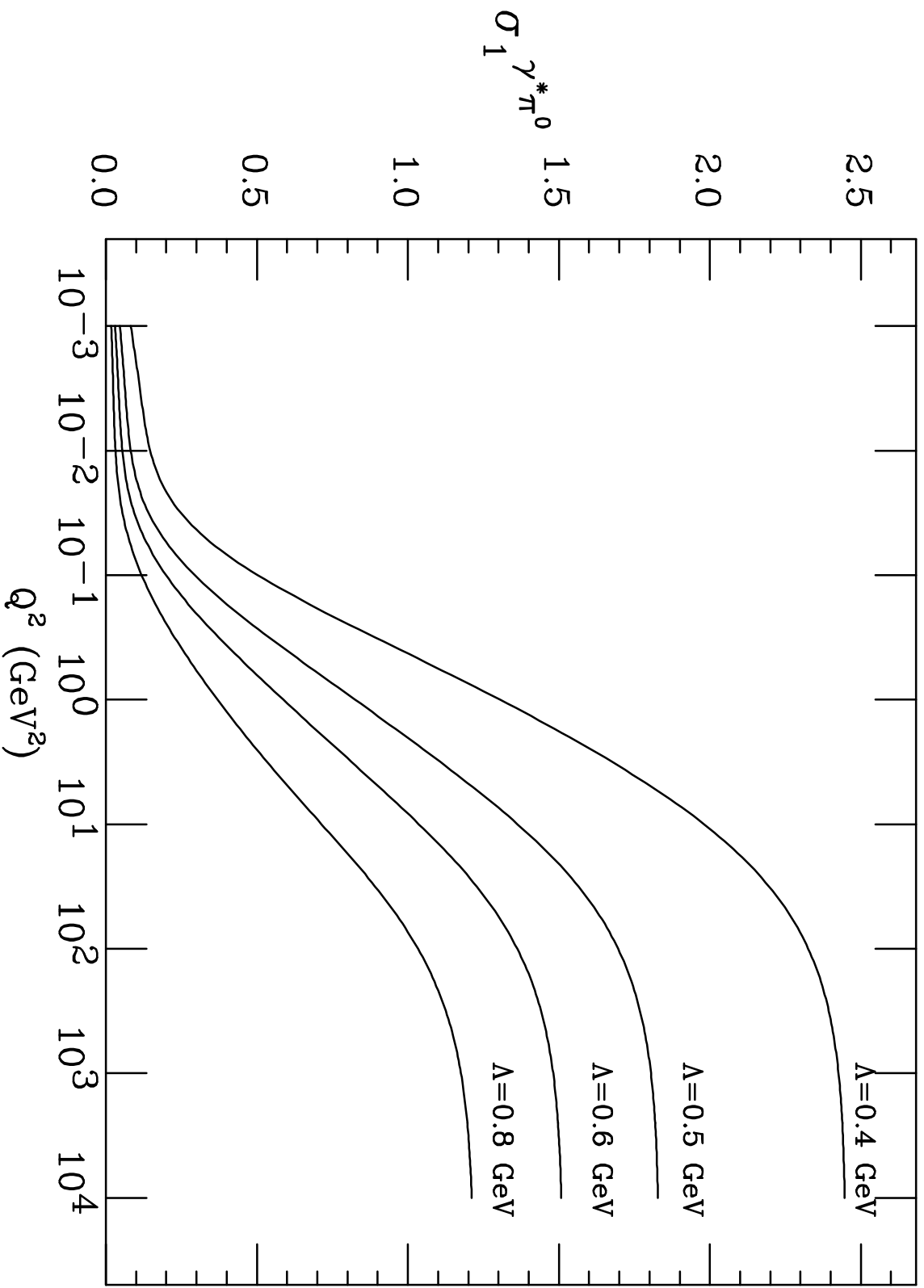
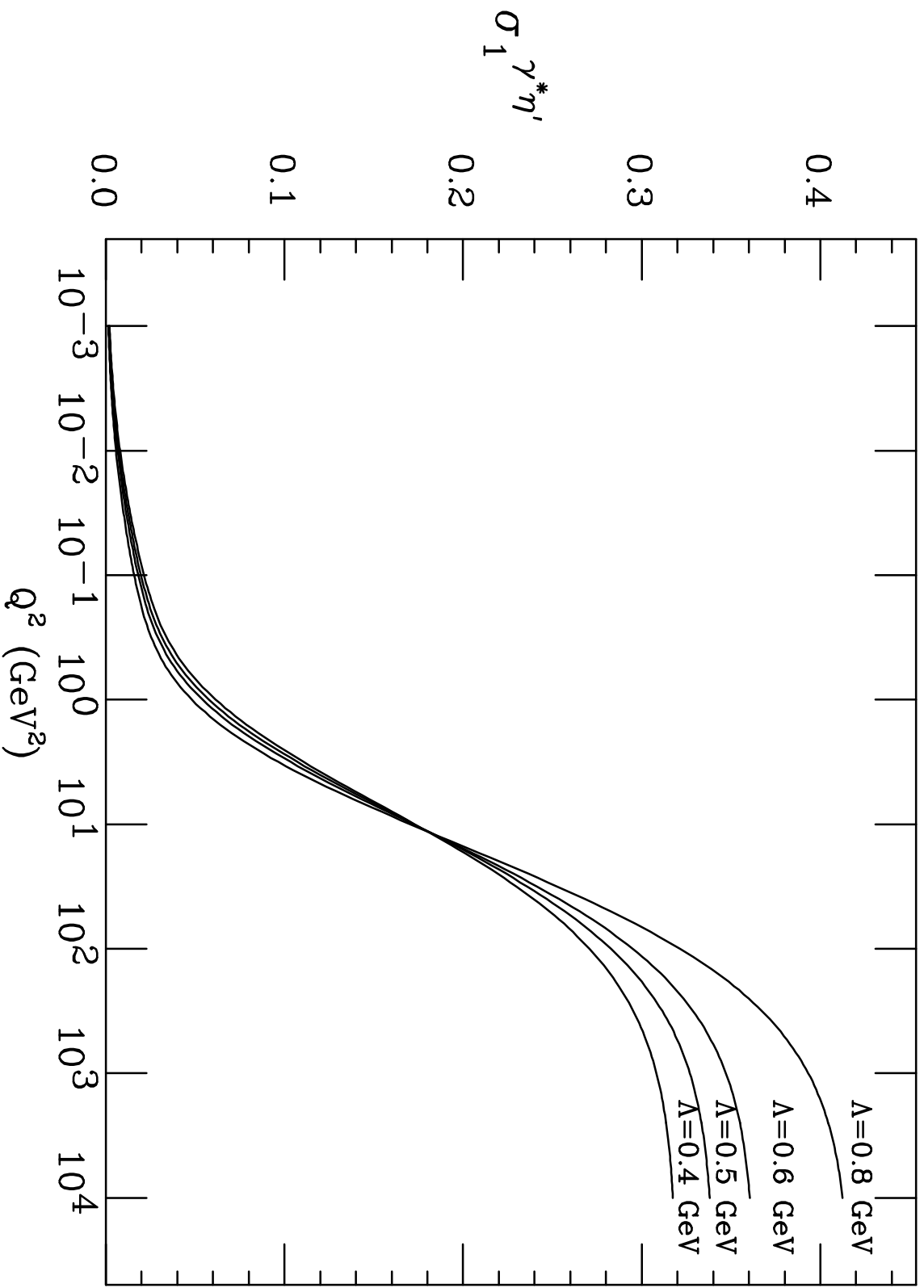
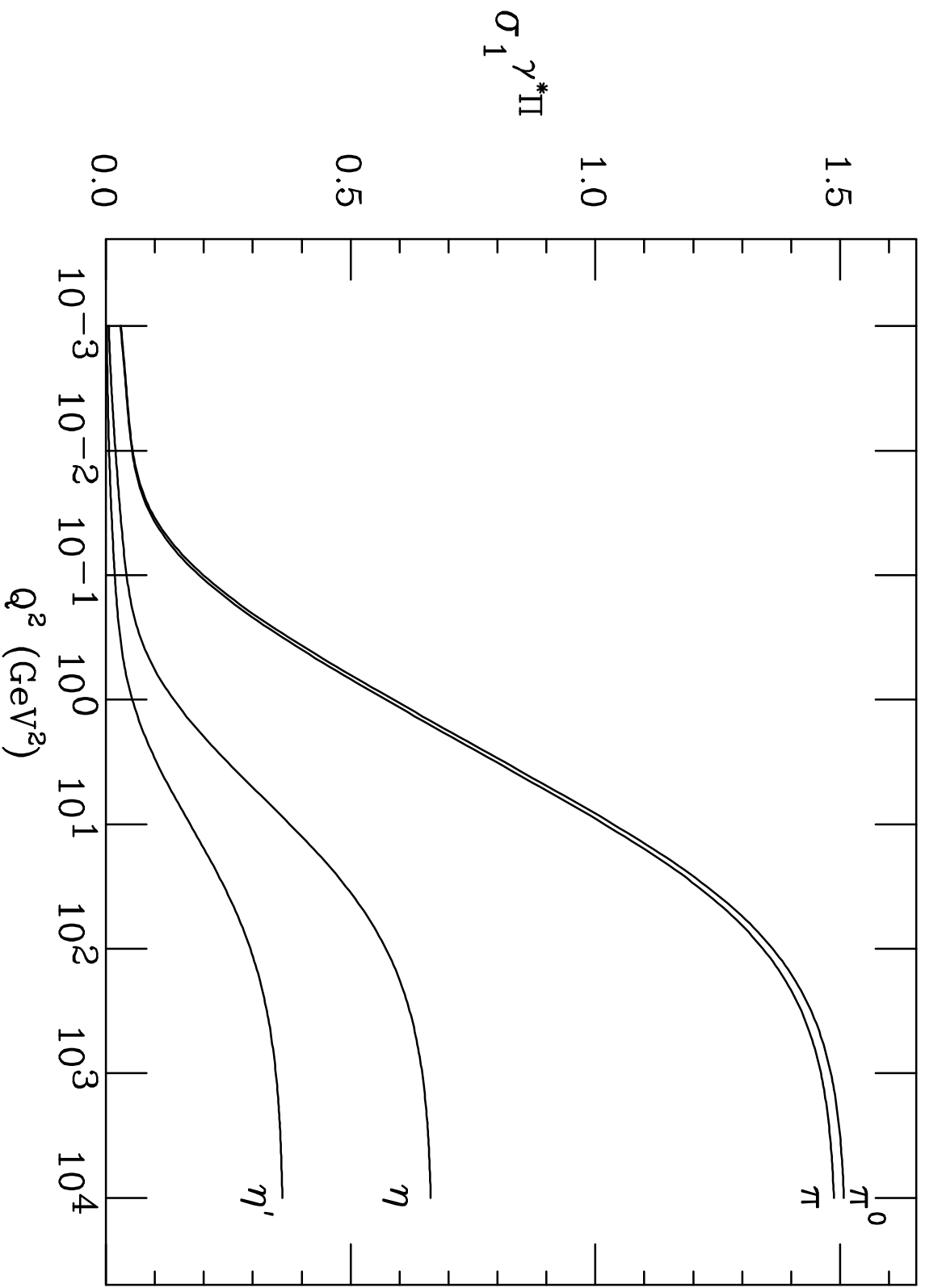
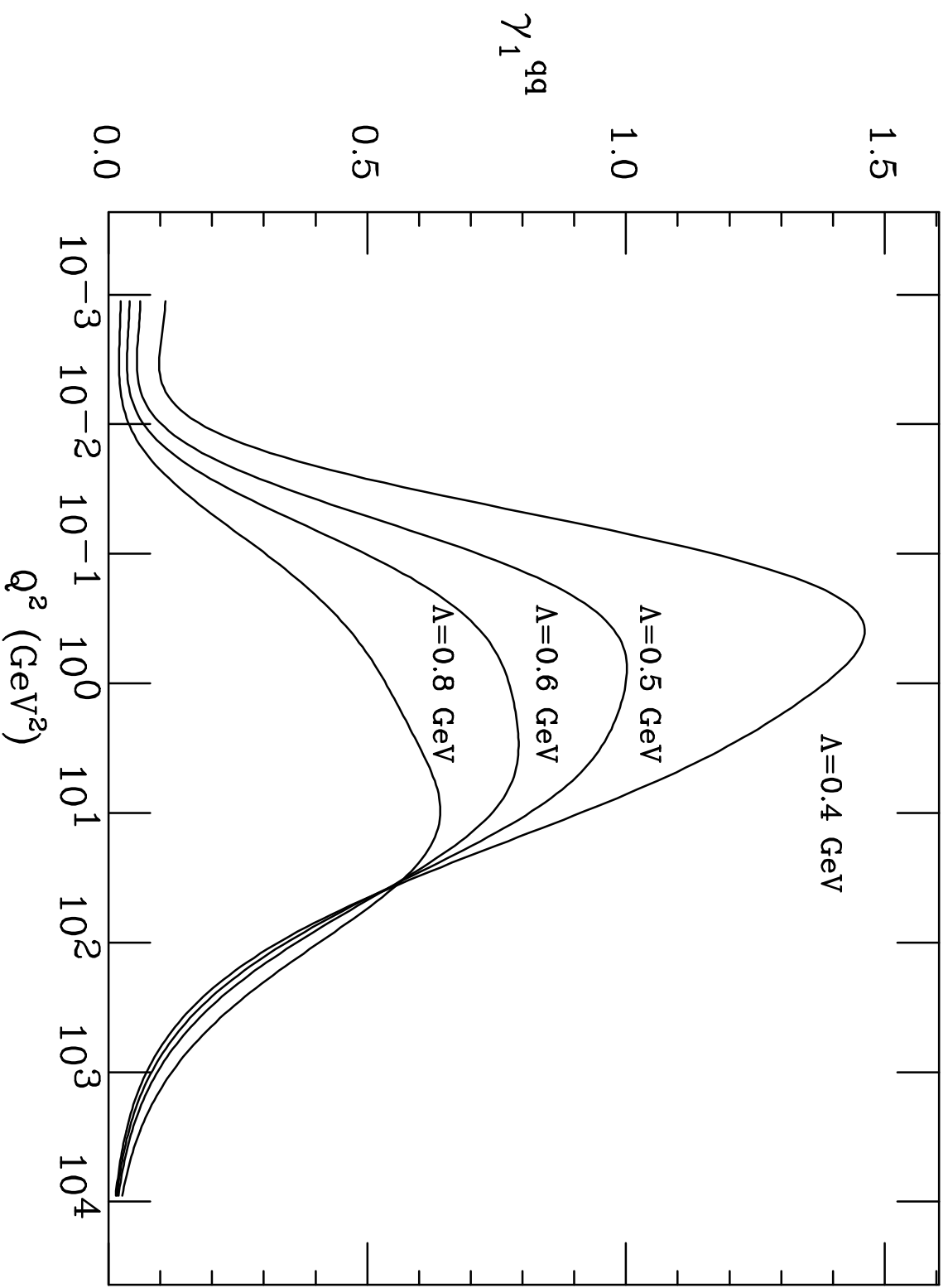
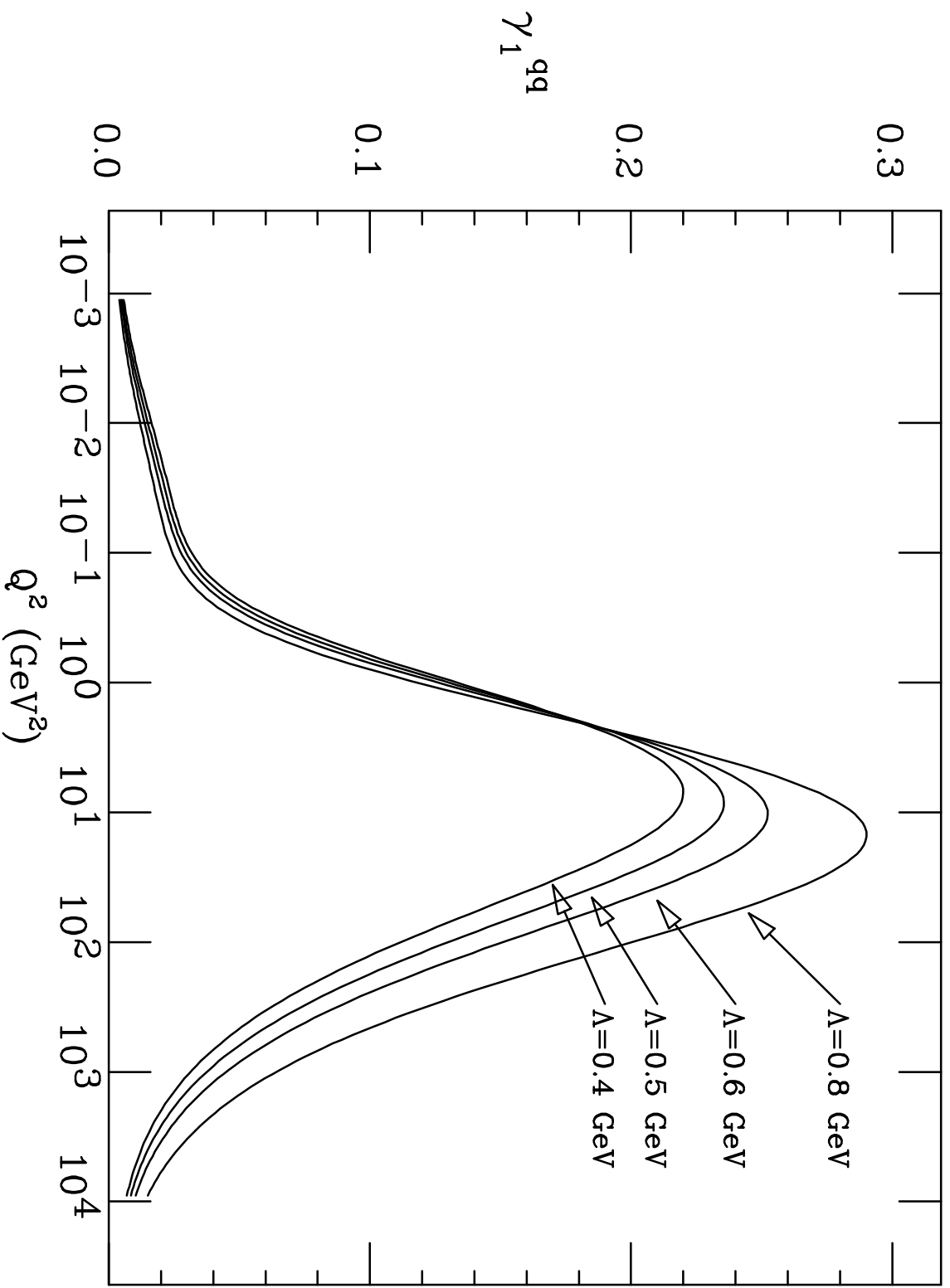
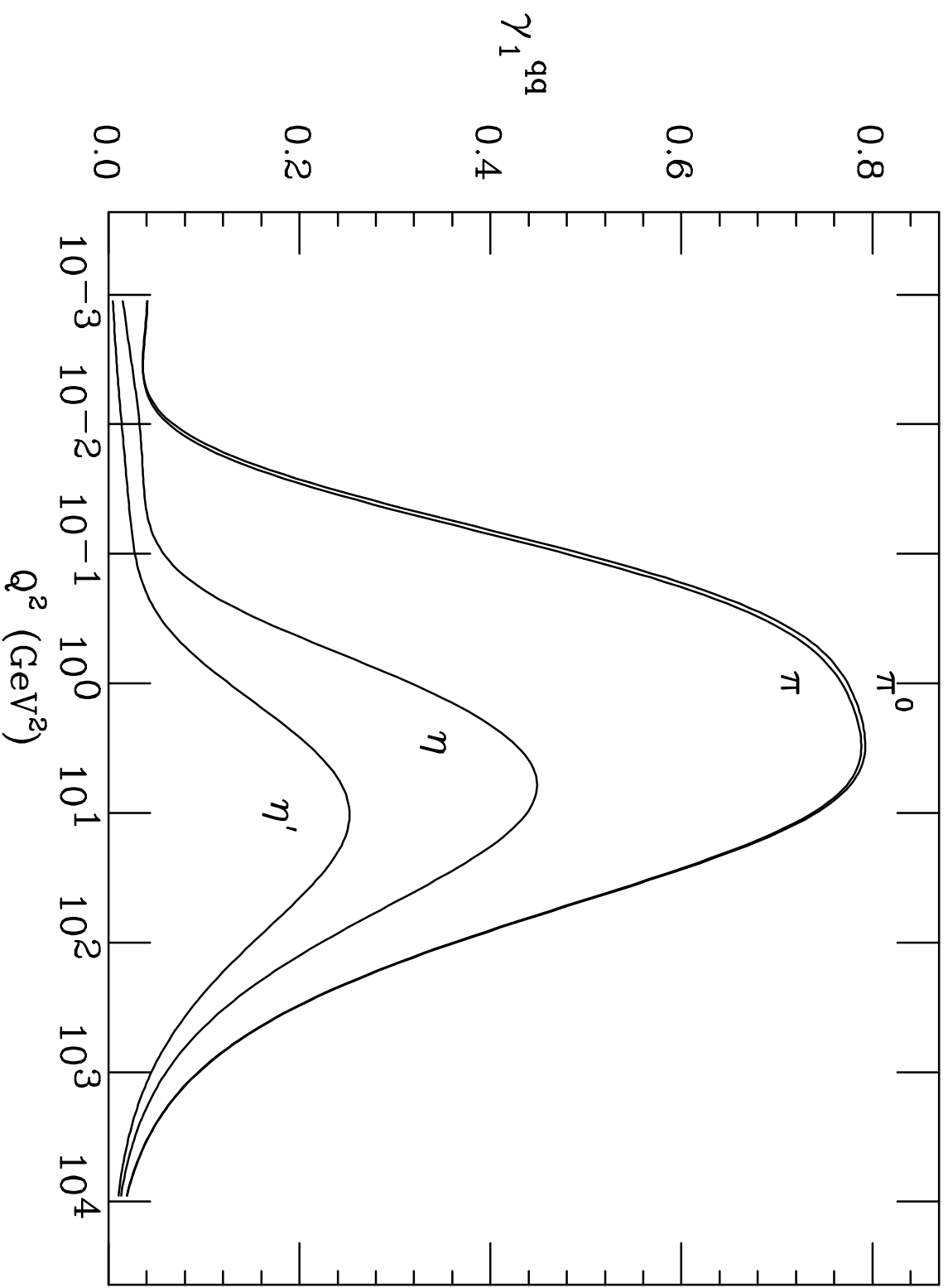
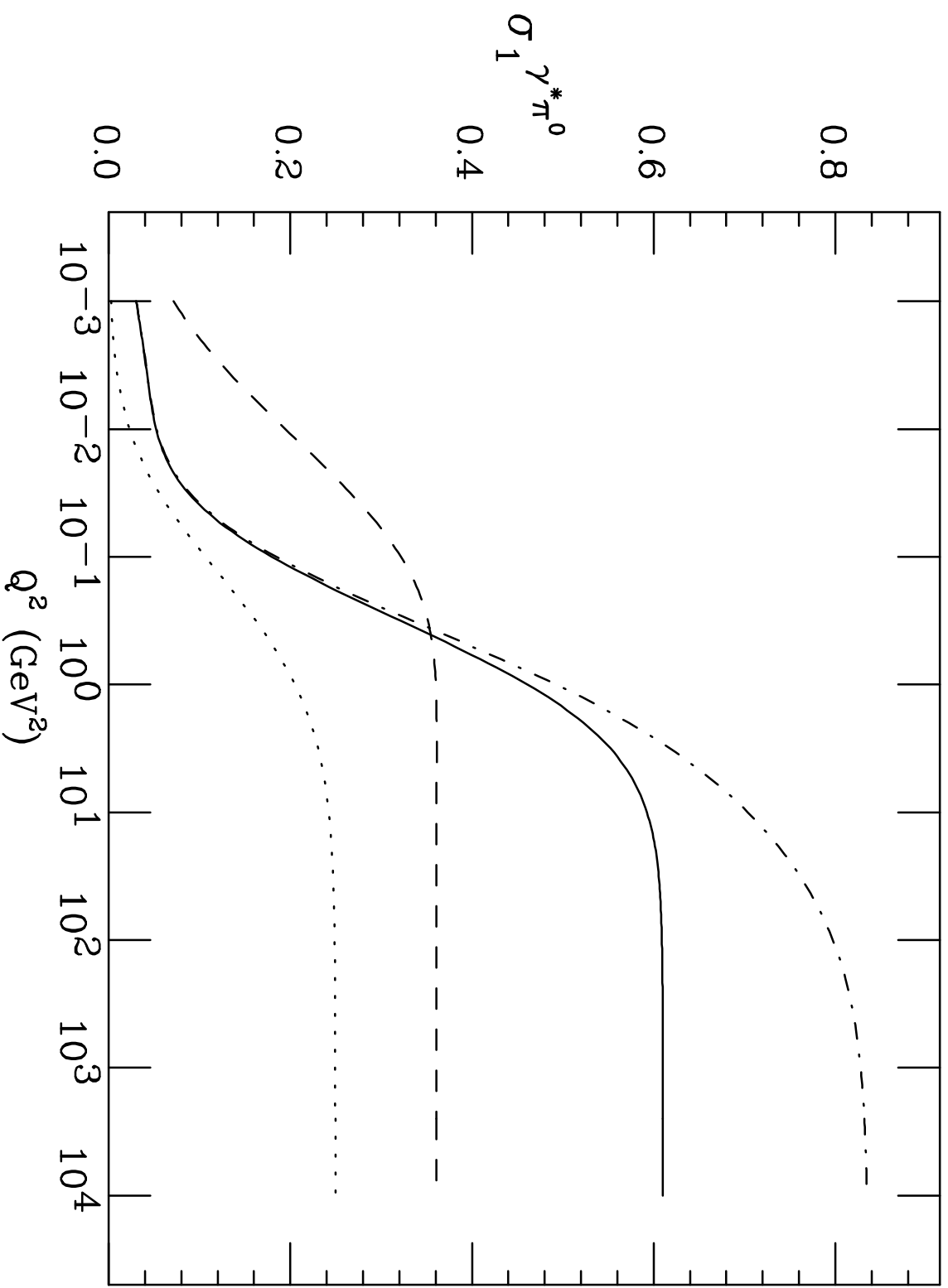
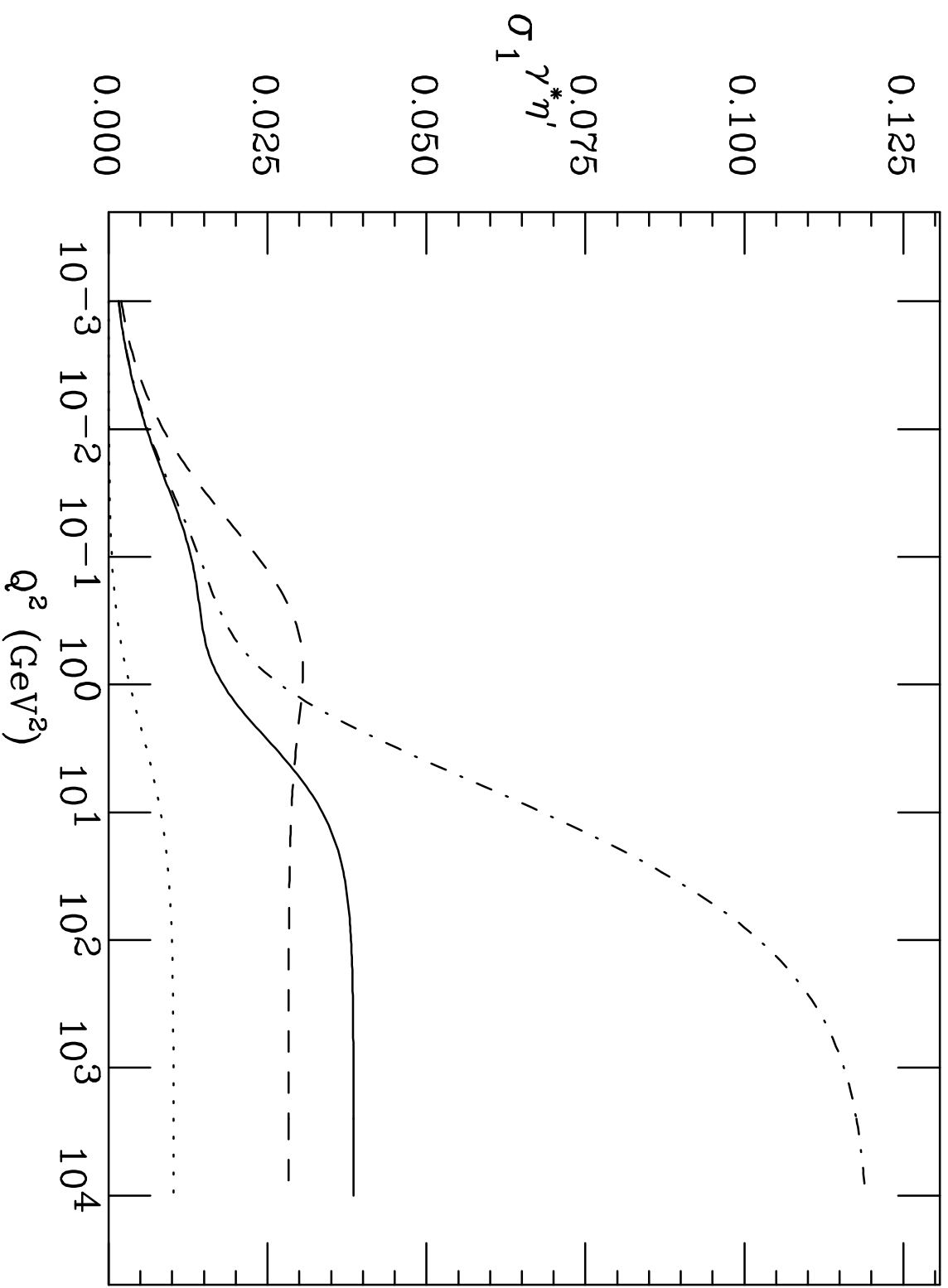
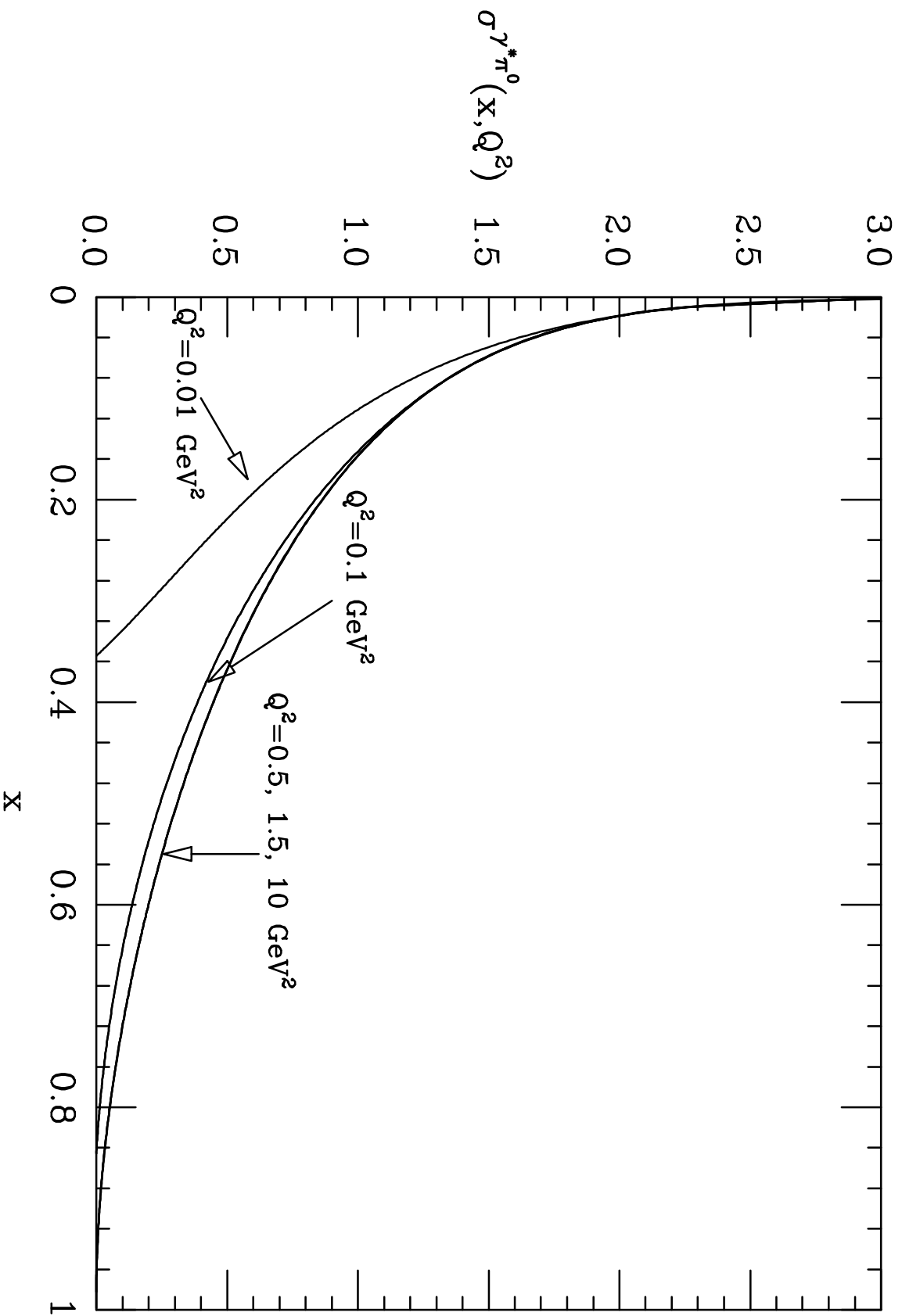
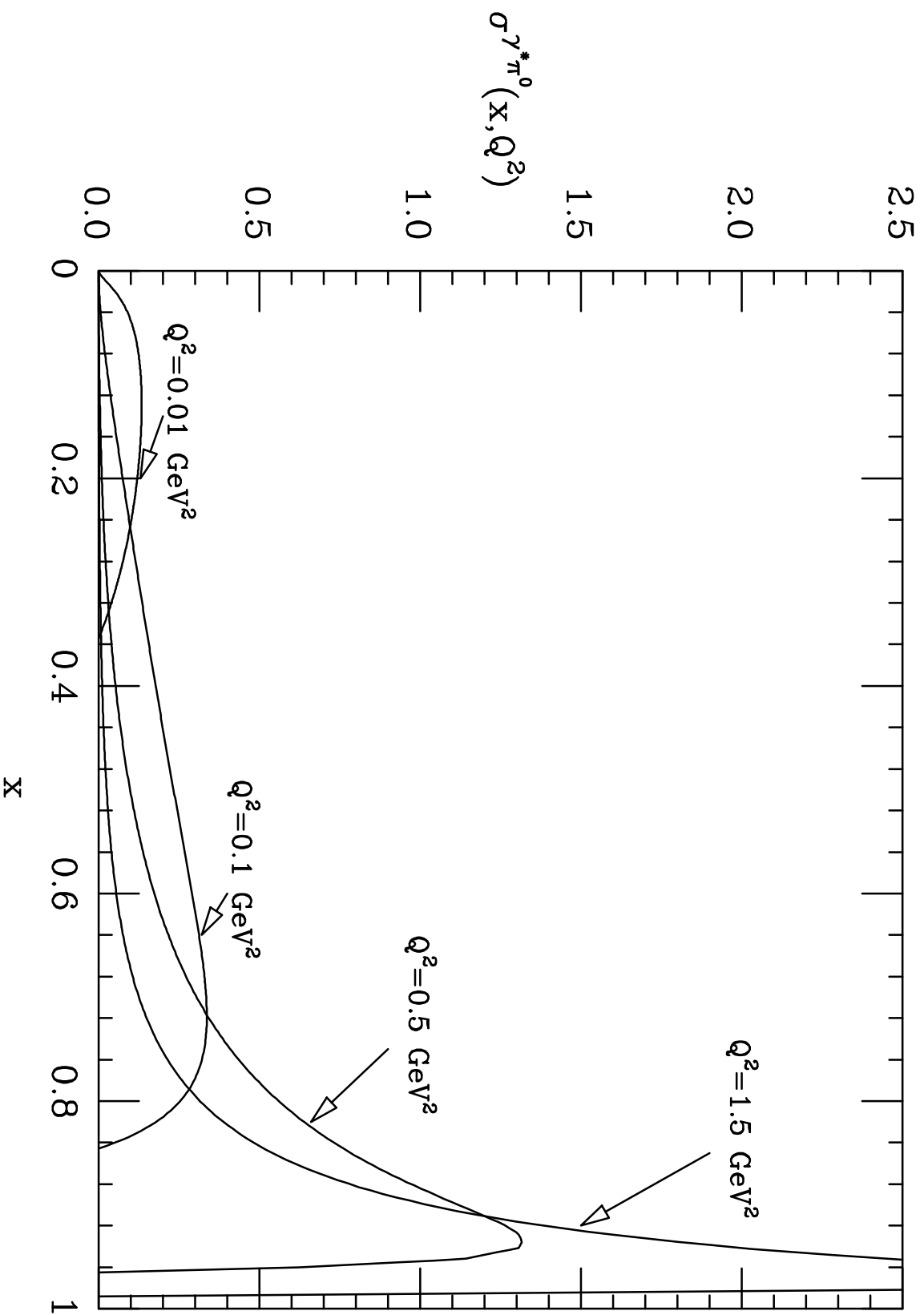
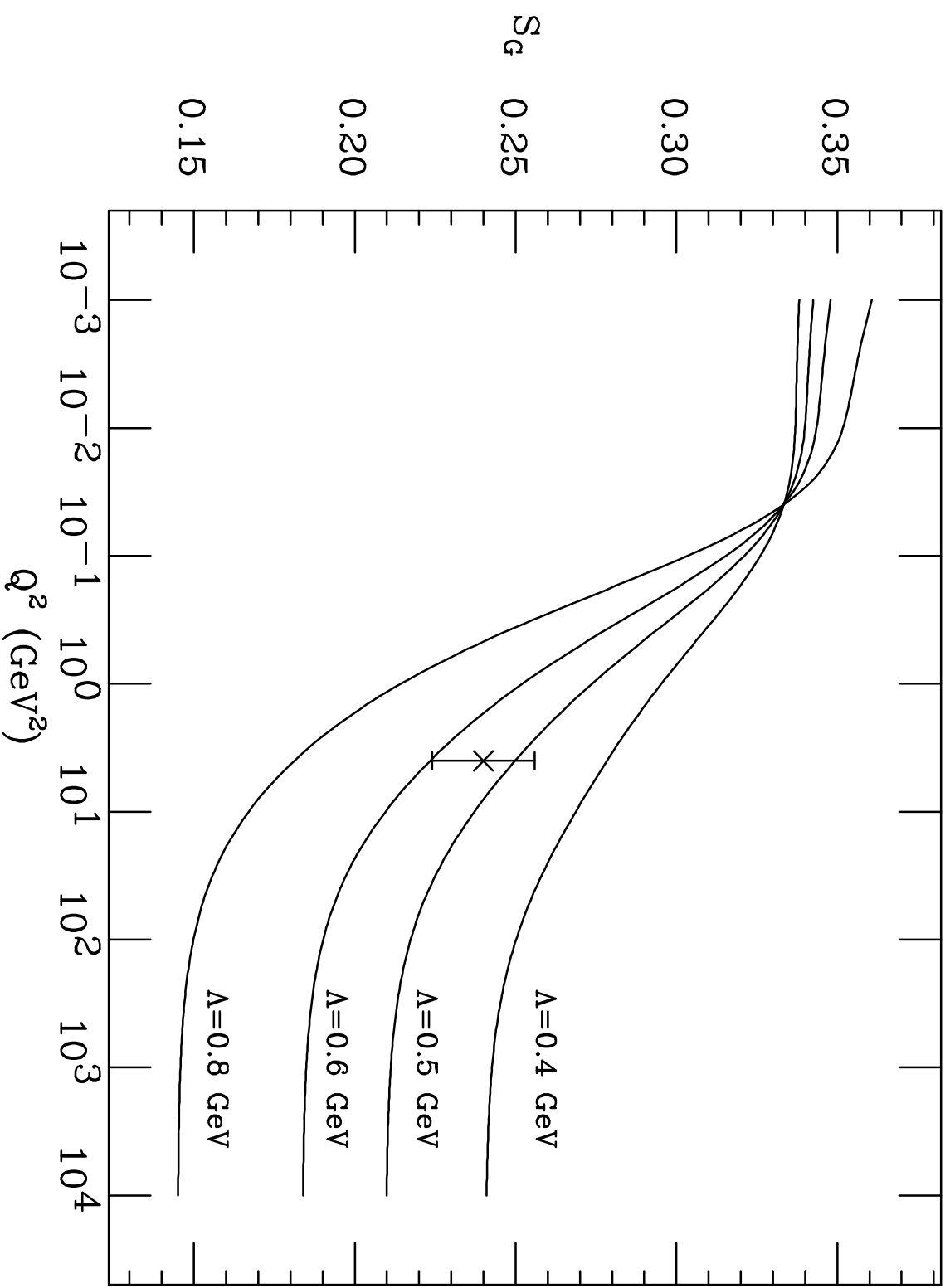
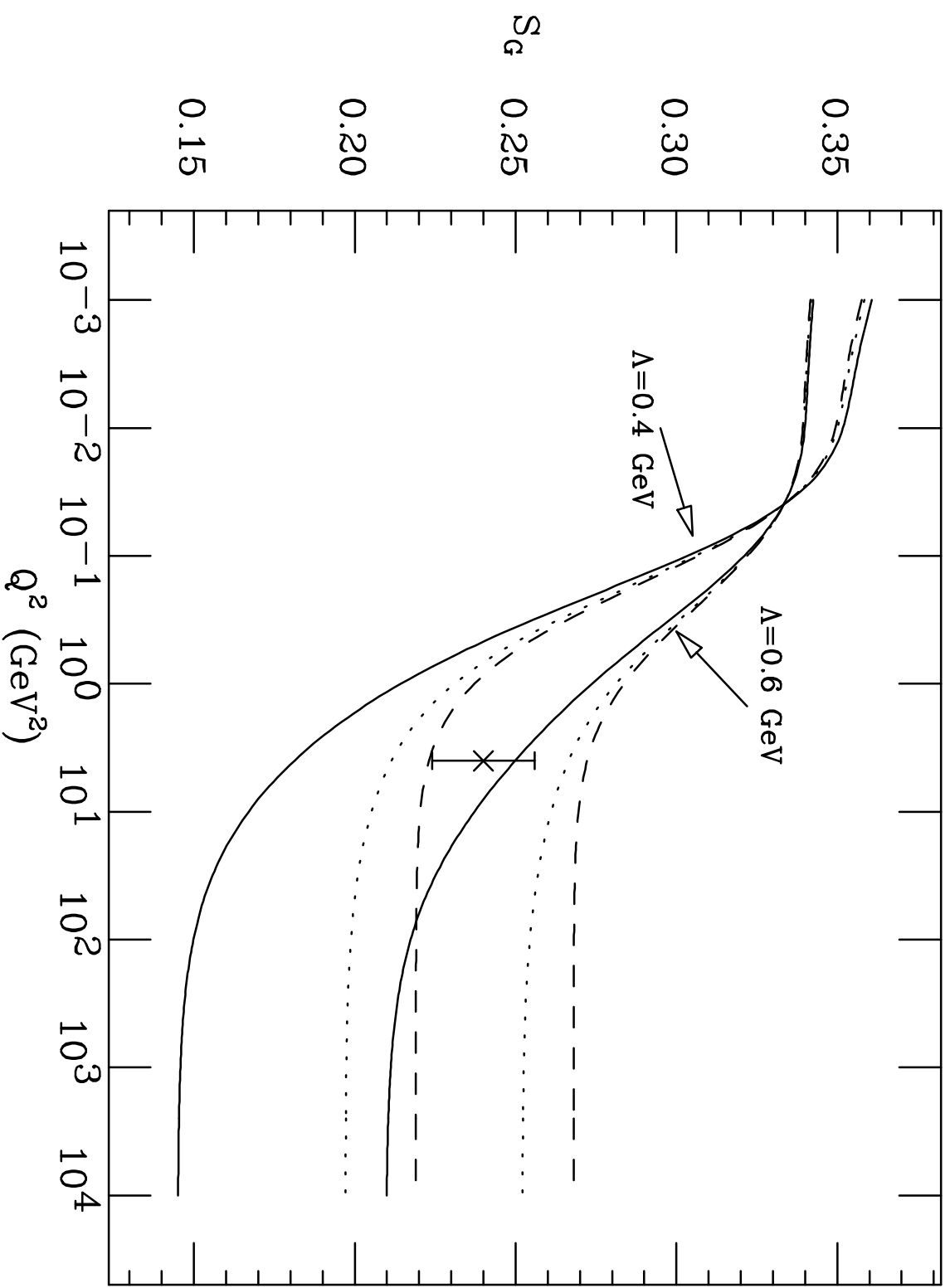
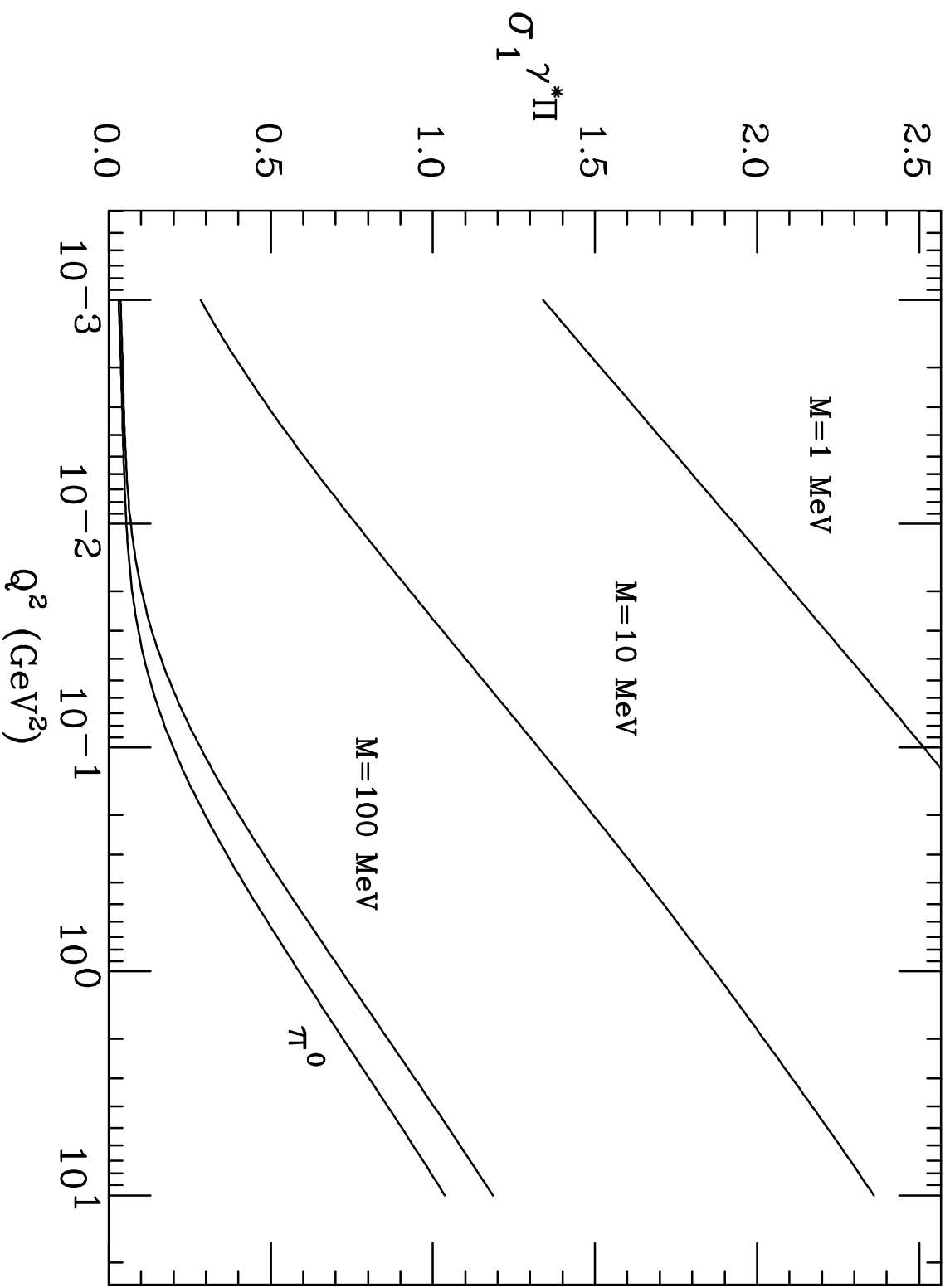
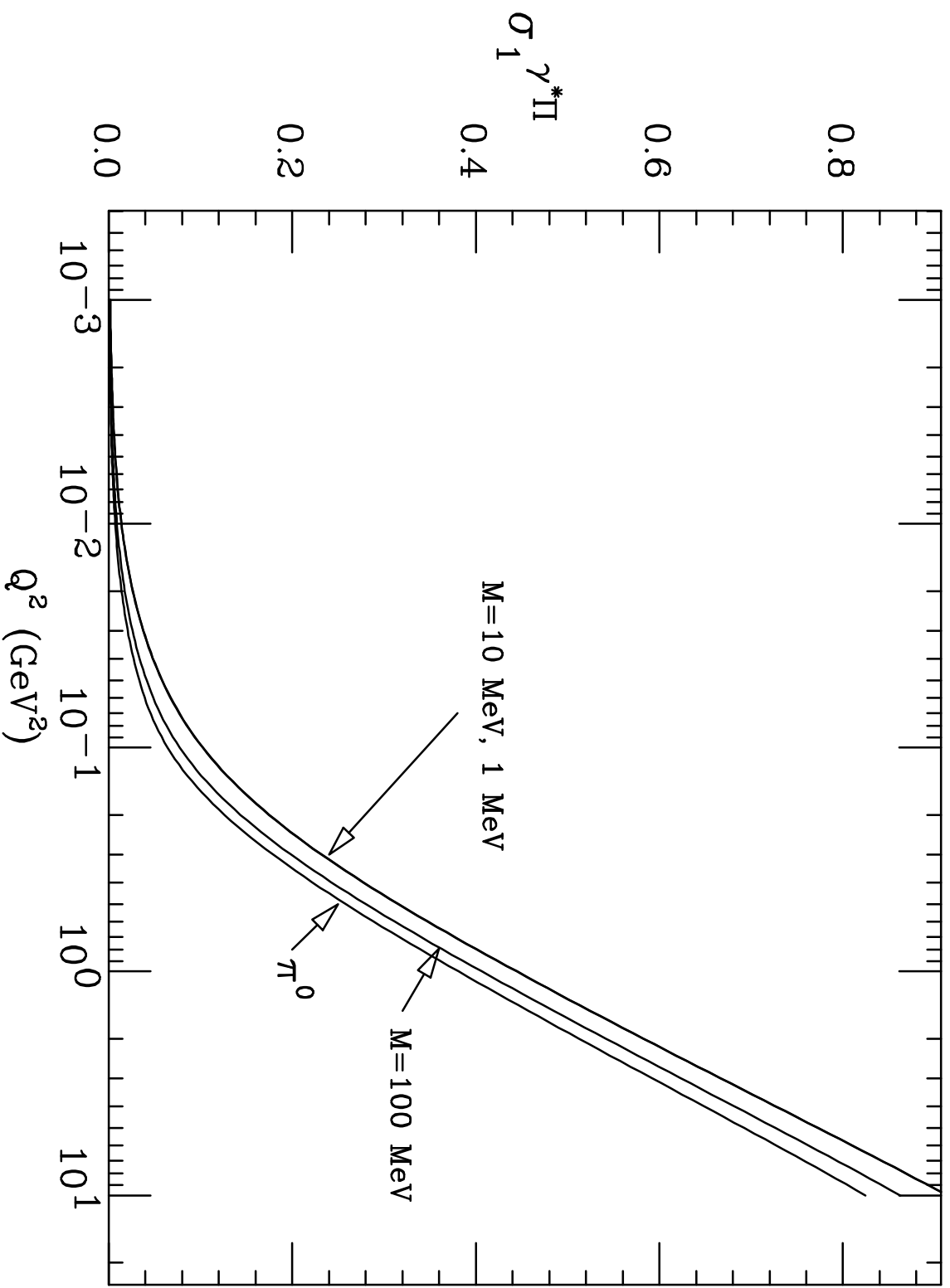
\bye